\documentclass[aps,pre,showpacs,floats,floats,twocolumn,superscriptaddress,floatfix,english]{revtex4-1}

\usepackage[T1]{fontenc}
\usepackage[latin9]{inputenc}
\usepackage{amsmath}
\usepackage{amssymb}
\usepackage{graphicx}

\usepackage{array}
\usepackage{multirow}
\usepackage{setspace}
\usepackage{subfigure}
\usepackage{caption}
\usepackage{color}
\usepackage{epstopdf}
\usepackage[normalem]{ulem}
\usepackage{float}
\usepackage{chemformula}
\usepackage{siunitx}

\begin{document}

\newtheorem{lemma}{Lemma}
\newtheorem{corollary}{Corollary}

\newcommand{\ju}[1]{\textcolor{red}{{\it#1}} }
\newcommand{\new}[1]{\textcolor{magenta}{{\it[#1]}} }
\newcommand{\bl}[1]{\textcolor{blue}{#1}}
\newcommand{\be}{\begin{equation}}
\newcommand{\ee}{\end{equation}}
\newcommand*\diff{\mathop{}\!\mathrm{d}}
\newcommand*\Diff[1]{\mathop{}\!\mathrm{d^#1}}
\providecommand{\tabularnewline}{\\}
\newcommand{\lyxdot}{.}
\newcommand{\tm}[1]{\textcolor{magenta}{#1}}
\newcommand{\tb}[1]{\textcolor{black}{#1}}
\newcommand{\tr}[1]{\textcolor{red}{#1}}
\newcommand{\tg}[1]{\textcolor{green}{#1}}
\def\D{\mathcal D}

\def\Hovmoller{H\"ovm\"oller }
\newcommand{\nordita}{Nordita, KTH Royal Institute of Technology and Stockholm University, Stockholm 10691, Sweden}
\def\D{\mathcal D}


\title{The combined effects of buoyancy, rotation, and shear {on phase boundary evolution}}

\author{S. Ravichandran}
\affiliation{\nordita}
\author{S. Toppaladoddi}
\affiliation{University of Oxford, Oxford OX1 4AL}
\author{J. S. Wettlaufer}
\affiliation{\nordita}
\affiliation{Yale University, New Haven, Connecticut 06520-8109, USA}

\makeatother

\begin{abstract}
We use well-resolved numerical simulations to study the combined effects of buoyancy, {pressure-driven} shear and rotation on the melt rate and morphology of a layer of pure solid overlying its liquid phase in three dimensions at a Rayleigh number {$Ra=1.25\times 10^5$}. {During thermal convection we find that the rate of melting of the solid phase varies non-monotonically with the strength of the imposed shear flow. In the absence of rotation, depending on whether buoyancy or shear {dominates the flow}, we observe either domes or ridges aligned in the direction of the shear flow respectively}.  Furthermore, we show that the geometry of the phase boundary has important effects on the {magnitude and evolution of the} heat flux in the liquid layer. In the presence of rotation, {the strength of which is} characterized by the Rossby number, $Ro$, we observe {that} for $Ro={O}(1)$ the mean flow in the interior is perpendicular to the direction of the {constant horizontal applied pressure gradient. 
As the magnitude of this pressure gradient increases,} the geometry of solid-liquid interface evolves from the voids characteristic of melting by rotating convection, to grooves oriented perpendicular or obliquely to the direction of the pressure gradient.
\end{abstract}

\maketitle



\section{Introduction\label{sec:Introduction}}

Fluid motions that accompany solid-liquid phase transitions are ubiquitous in both the natural and engineering environments \citep{epstein1983, glicksman1986, huppert1986, worster2000, meakin2009, hewitt2020}. Such fluid motions are either a result of the generation of unstable density gradients during solidification \citep{davis1984, dietsche1985, wettlaufer1997, worster1997, wykes2018}, or are externally imposed {and thereby influence} morphological and/or hydrodynamic instabilities in the system \citep{delves1968, delves1971, Gilpin1980, coriell1984, forth1989, feltham1999, neufeld2008, neufeld2008shear, Dallaston2015, ramudu2016, bushuk2019}. The interactions between fluid flows and phase-changing surfaces {underlie our understanding of, among other phenomena,} the evolution of Earth's cryosphere \citep{mcphee_book, hewitt2020}, defects in materials {castings} \citep{kurz}, and production of single crystals for silicon chips \citep{worster2000}. In this study, we focus on understanding the combined effects of buoyancy, rotation, and mean shear on the evolution of a phase boundary of a pure material.

The directional solidification of a pure melt is qualitatively different from that of a multi-component melt. However, there are qualitative similarities in the response of the two systems to externally imposed shear flows \citep{TW2019}. During the solidification of a multi-component melt, solute particles {(atoms, molecules or colloids)} are rejected into the bulk liquid, which can result in the constitutional supercooling of the liquid adjacent to the phase boundary \citep{worster2021}. This promotes ``morphological instability'', which leads to dendritic growth: the formation of a convoluted crystal matrix with a concentrated solution of solute particles trapped in the interstices \citep{worster2000}. This region is called a mushy layer. Depending on the equation of state of the liquid phase and the growth direction relative to the gravitational field, convection within (the ``mushy layer mode'') or adjacent to (the ``boundary layer mode'') {the mushy layer} may result \citep{worster1992}. Externally imposed flows can change the nature of these two modes.

Some of the earliest systematic investigations into the effects of a mean shear flow on the directional solidification of binary alloys are due to \citet{delves1968, delves1971} and \citet{coriell1984}. \citet{delves1968, delves1971} considered the effects of a parabolic flow on morphological instability using linear stability analysis in {two dimensions (2D)}. He found that the parabolic flow suppresses the instability, the degree of which depended on material properties. Similarly, \citet{coriell1984} studied the effects of a Couette flow on the morphological and thermo-solutal instabilities in {three dimensions (3D), and found that the latter are suppressed more than the former, which is due to the fact that the wavelength of morphological instability is several orders of magnitude smaller than the wavelength of thermo-solutal instability}. Furthermore, the use of a Couette flow as the base-state velocity profile in this case may only be locally valid, since  the momentum equations are governed by an asymptotic-suction boundary-layer profile \citep{drazin2004, TW2019}.

\citet{forth1989} {studied} the effects of an asymptotic-suction boundary-layer flow on morphological instability during the directional solidification of a binary alloy {using linear stability analysis in 3D. They found that under certain conditions the shear flow inhibits the morphological instability and may lead to the development of travelling waves along the phase boundary. However, the phase boundary was found to have negligible effects on the stability of the shear flow itself. }

In the directional solidification of {mushy layers}, the fate of any incipient perturbation introduced at the mush-melt interface naturally depends on the interaction between the mushy-- and boundary--layer modes of convective motion. In order to understand the evolution of such perturbations, \citet[][see also the corrigendum in \cite{neufeld2006}]{feltham1999} considered the effects of flows of inviscid and viscous melts on the mush-melt interface using linear stability analysis {in 2D}. Neglecting buoyancy, they found that the forced flows over a corrugated interface introduce pressure perturbations at the phase boundary, which in turn drive fluid motions in the mushy layer. These fluid motions, in certain cases, were found to amplify the interfacial perturbations. However, the heat flux from the melt was found to have negligible effects on this amplification, and was found to only generate travelling waves at the interface.

\citet{neufeld2008, neufeld2008shear} used a combination of linear stability analysis {in 3D} and experiments to examine the effects of a shear flow on the convective motions in the mushy layer and bulk melt. Their main findings are: (i) below a critical strength of the imposed shear flow, the convective motions are moderately suppressed; (ii) above this critical strength, the stability of these convective motions decreases monotonically with increasing shear-flow strength; and (iii) for a sufficiently strong shear flow, quasi-two dimensional striations of zero volume fraction of solid are formed perpendicular to the shear-flow direction at the mush-melt interface. These striations form due to local dissolution and growth of the mushy layer, which result from the interactions between the shear and convective motions within and outside the mushy layer.

In addition to shear flows, several studies have explored the effects of rotation on the stability of the mush-melt interface and the convective motions inside and outside of the mushy layer. 
For example, \citet{sample1982, sample1984} {experimentally} explored the effects of rotation and rotational-and-precessional motions {during} directional solidification of \ch{NH4Cl-H2O} and \ch{Pb-Sn} systems, respectively. Their key findings are: (i) channels in the mushy layer through which solute is expelled into the bulk melt originate close to the mush-melt interface due to perturbations in the solutal boundary layer; (ii) when the mould rotates about an angle to the vertical (\ang{20} - \ang{30}) at low speeds ($< 5$ rpm), the induced {interfacial} shear flow arrests the growth of the perturbations, thereby inhibiting the formation of channels; and (iii) rotation about the vertical axis has very little effect on the evolution of the perturbations. In a {related} study, \citet{lu1997} used linear stability analysis to show that the rotation about a vertical axis has an appreciable stabilizing effect only for very large rotation rates ($\sim 10^5$ rpm).

Motivated by the experiments of \cite{sample1984}, Chung \& Chen used linear stability analysis to study the detailed effects of inclined rotation and precessional motions on the stability of {the} mushy layer \citep{chung2000}, {the} mush-melt interface and the bulk melt \citep{chung2003}. {They found that} (i) rotation of the system about an inclined axis induces flow parallel to the mush-melt interface in both the mushy layer and the bulk melt. However, the maximum velocity in the bulk is much larger than the maximum velocity in the mushy layer \citep{chung2000}{;} (ii) the induced velocity in the melt increases with increasing angle of inclination and decreases with rotation rate, but the induced velocity in the mushy layer is only sensitive to the inclination angle and increases with it \citep{chung2000}{;} (iii) the stabilization of the mushy layer is largely due to the reduction in buoyancy perpendicular to the mush-melt interface \citep{chung2000}{;} (iv) when there is only precessional motion, the most unstable mode of instability in the mushy layer is in the direction perpendicular to the mush-melt interface. Introducing rotation allows the induced flow to explore all directions equally, thereby equally stabilizing all modes \citep{chung2000}{;} (v) five competing {mechanisms -- the (stabilising) reduction of buoyancy and rotation normal to the interface, the (destabilising) component of gravity parallel to the interface, and the (stabilising or destabilising) flow induced by inclination and precession --} determine the stability of the bulk melt and the interface \citep{chung2003}.

There have been relatively few studies on the effects of buoyancy, mean shear, and rotation on directional solidification of pure melts. Using linear and weakly nonlinear stability analyses and experiments, \citet{davis1984} explored the different patterns on the phase boundary that result due to {thermal} convection. These patterns and the transition between them were further explored experimentally by \citet{dietsche1985}. Recent studies have focussed on the effects of high Rayleigh-number convection on the phase boundary \citep{esfahani2018, favier2019, purseed2020}. A detailed summary of these studies can be found in \citet{toppaladoddi2020}.


\citet{hirata1979a, hirata1979} and \citet{Gilpin1980} experimentally studied the effects of laminar and turbulent boundary-layer flows on the evolution of phase boundaries between ice and water. \citet{Gilpin1980} observed that {when the conductive heat flux through the ice is less than approximately twice the water-to-ice flux}, an initially introduced perturbation at the interface grew and advected downstream, leading to the formation of ``rippled'' surfaces. The heat transfer rate increased by about 30-60\% over these corrugated surfaces when compared with a planar interface. These observations were attributed solely to mean shear by \citet{Gilpin1980}, but a recent reanalysis of their data showed that there were substantial buoyancy effects due to anomalous behaviour of water at 4 $^\circ$C in their experiments \citep{TW2019}. Furthermore, a linear stability analysis of the Rayleigh-B\'enard-Couette flow over a phase boundary showed that mean shear has a stabilizing effect and buoyancy has a destabilizing effect on the phase boundary \citep{TW2019}. More recently, the interplay between mean shear, buoyancy, and phase boundaries has been further explored in {2D} \citep{toppaladoddi2020} and {3D} \citep{couston2020} using direct numerical simulations.

\citet{ravichandran2021} explored the combined effects of rotation and thermal convection on the evolution of a phase boundary {in 3D}. They found that the rotation-dominated convection of a pure liquid, which takes the form of columnar vortices \citep{rossby1969}, melts voids into its {overlying} solid phase. They showed that the feedback of the melting on the convective flow can arrest the horizontal motion of flow structures like the peripheral streaming flow in wall-bounded rotating convection, and hypothesized that the columnar vortices may also be pinned to the locations of the voids. Furthermore, their study revealed that the degree of interfacial roughness co-varies with the heat flux.

Here, we use direct numerical simulations to study the melting of the solid phase of a pure substance driven by thermal convection of its melt, subject to a constant {applied pressure gradient} in the horizontal direction and rotation about the vertical axis. The combined effects of shear, rotation, and buoyancy on the evolution of phase boundaries are important in several geophysical settings \citep{wells2008, ramudu2016, hewitt2020}, and provide important motivations for this study.

The {remainder} of the paper is organised as follows. In \S \ref{sec:Setup}, we formulate the problem and present the equations of motion along with the initial and boundary conditions. We briefly discuss the numerical method used to solve the equations of motion in \S \ref{sec:num-method}. We then discuss results from simulations of non-rotating flows in \S \ref{subsec:Zero-rotation} and of rotation-influenced flows in \S \ref{subsec:Finite-rotation}. Finally, we conclude with a summary of our observations and suggestions for future work in \S \ref{sec:Conclusions}.


\section{Problem formulation and governing equations \label{sec:Setup}}

\begin{figure}
\noindent \centering{}\includegraphics[trim = 100 30 100 150, width=0.75\columnwidth]{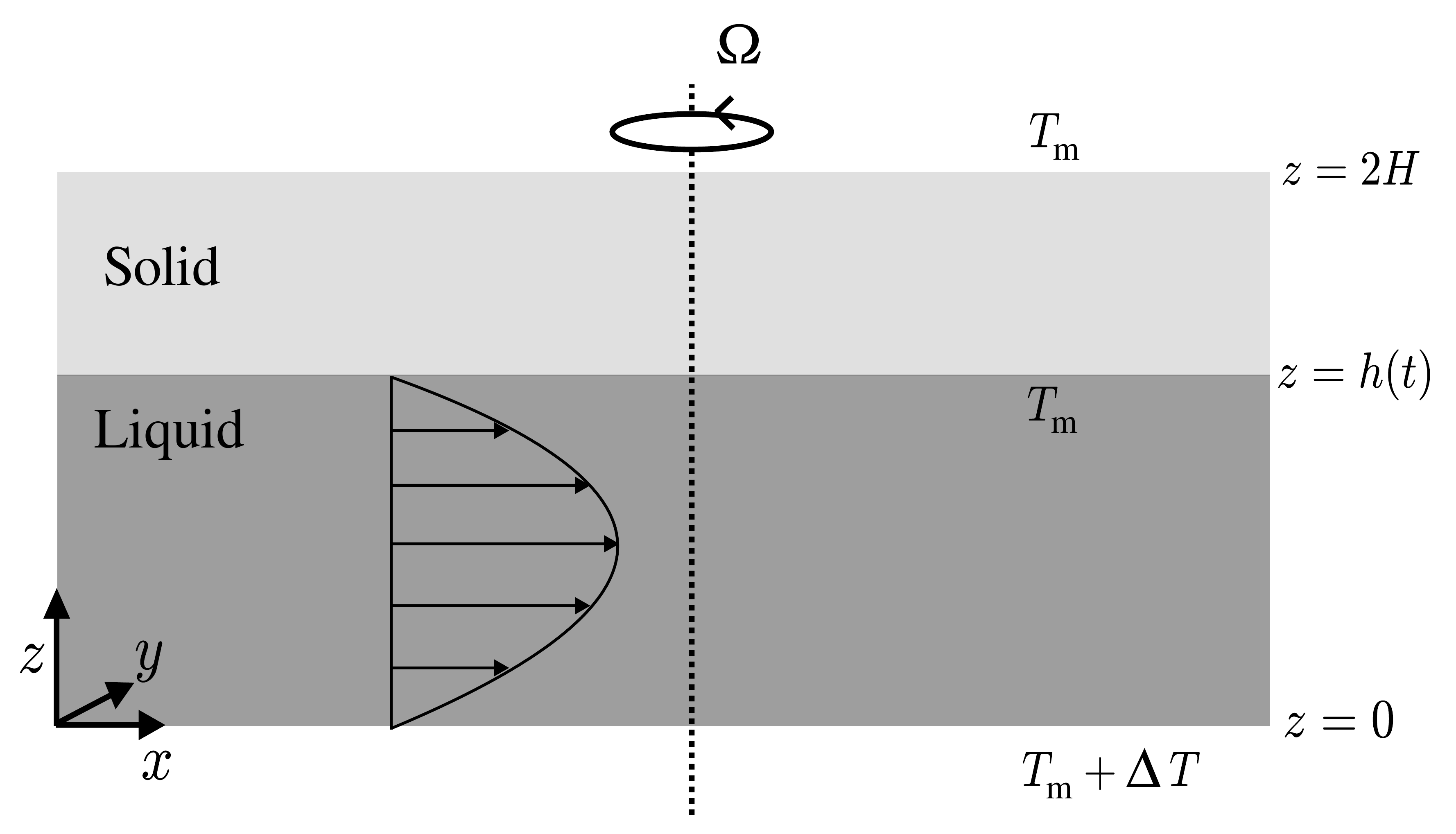}
\caption{\label{fig:schematic} A vertical cross-section of the geometry considered.
The size of the domain is $L \times L\times 2H$, with the aspect ratio $A=L/2H$. The bottom surface is maintained at a higher temperature than the interface. No-slip and no penetration conditions for the fluid velocity are imposed at the bottom boundary and the evolving interface, and periodic boundary conditions are imposed in the horizontal.}
\end{figure}

In this study, we consider a cuboid of dimensions of $L \times L \times 2H$. The aspect ratio of the cell is defined as $A = L/2H$ and is fixed at either $4$ or $8$ in the simulations reported here. Figure \ref{fig:schematic} shows a vertical cross section of the domain. 
At the initial instant, the planar phase boundary is located at {$z = h_0$} and the fluid is contained in the region {$0 \le z \le h_0$}.  By definition, the temperature at the phase boundary is $T = T_m$, and that at the top plate, $z = 2H$, is maintained at $T = T_m$.
The bottom plate located at $z = 0$ is maintained at a temperature above the melting point, $T = T_m + \Delta T$. We note that because of the isothermal temperature field in the solid phase, it completely melts before a stationary state is reached. Thus, we focus solely on the system dynamics until the solid phase vanishes. Periodic boundary conditions are imposed in the horizontal directions. 

The equations of motion in the different regions of the domain {are as follows.}
\subsection{Liquid}
The conservation of {momentum, mass} and heat are given by
\be
\frac{\partial \boldsymbol{u}}{\partial t} +  \boldsymbol{u} \cdot \nabla  \boldsymbol{u} = -\frac{1}{\rho_0} \, \nabla p + g \, \alpha \, \left(T-T_m\right) \, \boldsymbol{\hat{z}} - 2 \, \Omega \boldsymbol{\hat{z}} \times \boldsymbol{u} + \nu \, \nabla^2  \boldsymbol{u} + F_p \, \boldsymbol{\hat{x}},
\label{eqn:NS_liquid}
\ee
\be
\nabla \cdot \boldsymbol{u} = 0, \text{{and}}
\label{eqn:mass}
\ee
\be
\frac{\partial T}{\partial t} +  \boldsymbol{u} \cdot \nabla  T = \kappa \, \nabla^2  T,
\label{eqn:heat_liquid}
\ee
respectively. Here, $\boldsymbol{u}(\boldsymbol{x},t) = (u, v, w)$ is the three-dimensional velocity field; $\rho_0$ is the reference density; $p(\boldsymbol{x},t)$ is the pressure field; $g$ is acceleration due to gravity; $\alpha$ is the thermal expansion coefficient; $T(\boldsymbol{x},t)$ is the temperature field; {the constant pressure gradient is applied as a horizon{t}al body force per unit mass}, $F_p$; $\boldsymbol{\hat{x}}$ and $\boldsymbol{\hat{z}}$ are the unit vectors along the $x$ and $z$ axes, respectively; $\Omega$ is the constant speed of rotation; $\nu$ is the kinematic viscosity; and $\kappa$ is the thermal diffusivity.

\subsection{Solid}
The {entire} solid phase is isothermal because of the temperature boundary condition at the upper boundary. Hence, the temperature field in the solid phase is given by
\be
T(\boldsymbol{x},t) = T_m.
\ee

\subsection{Evolution of the phase boundary}
The location of the phase boundary is determined using the Stefan condition, which is a statement of energy balance \citep{worster2000}:
\be
\rho_0 \, L_s \, v_n =  - \boldsymbol{n} \cdot \boldsymbol{q_l} \hspace{0.2cm} \text{at} \hspace{0.1cm} z = h(x,y,t).
\label{eqn:stefan}
\ee
Here, $L_s$ is the latent heat of fusion; $v_n$ is the normal component of growth rate of the solid phase; $\boldsymbol{n}$ is the unit normal pointing into the liquid; and $\boldsymbol{q_l}$ is the heat flux towards the phase boundary from the liquid.

\subsection{Boundary conditions}
We impose Dirichlet conditions on temperature at the bottom and top boundaries of the domain:
\be
T(x, y, z=0,t) = T_m + \Delta T \quad \mbox{and} \quad T(x,y,z = H,t) = T_m,
\ee
where $\Delta T > 0$. {The temperature at the phase boundary} is the bulk melting temperature:
\be
T(x,y,z=h,t) = T_m.
\ee

We impose no-slip and no-penetration conditions on the flow velocity at the lower boundary and the phase boundary, given by
\be
u(x,y,z = 0, t) = v(x,y,z=0,t) = w(x,y,z = 0, t) = 0;
\ee
\be
\boldsymbol{u} \cdot \boldsymbol{n} = \boldsymbol{u} \cdot \boldsymbol{t_1} = \boldsymbol{u} \cdot \boldsymbol{t_2} = 0 \quad \mbox{at} \quad z = h(x,y,t),
\ee
respectively, where $\boldsymbol{t_1}$ and $\boldsymbol{t_2}$ are the $x$ and $y$ projections of the unit tangents at the phase boundary. {For simplicity, we assume that the reference densities of the solid and liquid phases are equal to $\rho_0$. Hence, there is no additional normal velocity induced at the phase boundary due to density difference between the phases}. We also impose periodic boundary conditions for the temperature and velocity fields at $x = 0$ and $x = L$ and $y = 0$ and $y=L$.

\subsection{Initial conditions}
At $t=0$, the temperature of the liquid layer is maintained at $T(\boldsymbol{x},t=0) = T_m$, and the velocity field is $\boldsymbol{u}(\boldsymbol{x},t=0) = \boldsymbol{0}$. The shear flow in the horizontal direction is driven by the body force starting at $t=0$.


\subsection{Non-dimensional equations of motion}
To non-dimensionalise the equations of motion and the associated boundary conditions, we choose $H$ as the length scale{;} $U_0 = \sqrt{g \, \alpha \, \Delta T \, H}$ as the velocity scale{;} $t_0 = H/U_0$ as the time scale{;} $p_0 = \rho_0 \, U_0^2$ and the pressure scale{;} and $\Delta T$ as the temperature scale. Using these in equations \ref{eqn:NS_liquid} -- \ref{eqn:heat_liquid} and \ref {eqn:stefan} and maintaining pre-scaled notation, we have 
\be
\frac{\partial \boldsymbol{u}}{\partial t} +  \boldsymbol{u} \cdot \nabla  \boldsymbol{u} = -\nabla p + \sqrt{\frac{Pr}{Ra}} \, \nabla^2 \boldsymbol{u} - \frac{2}{Ro} \, \boldsymbol{\hat{z}} \times \boldsymbol{u} + \theta \, \boldsymbol{\hat{z}} + {\frac{1}{Ri_b}} \, \boldsymbol{\hat{x}},
\label{eqn:NS_liquid_scaled}
\ee
\be
\nabla \cdot \boldsymbol{u} = 0,
\label{eqn:mass_scaled}
\ee
\be
\frac{\partial \theta}{\partial t} +  \boldsymbol{u} \cdot \nabla  \theta = \sqrt{\frac{1}{Pr \cdot Ra}} \, \nabla^2  \theta \text{{, and}}
\label{eqn:heat_liquid_scaled}
\ee
\be
v_n = \sqrt{\frac{1}{Pr \cdot Ra \cdot St^2}} \, \boldsymbol{n} \cdot \nabla \theta \hspace{0.2cm} \text{at} \hspace{0.1cm} z = h(x,y,t),
\label{eqn:stefan_scaled}
\ee
where {the dimensionless temperature,}
\be
\theta = \frac{T-T_m}{\Delta T},
\ee
{is defined such that $\theta=0$ at the phase boundary, and $\theta=1$ at the heated lower boundary.}
The five dimensionless parameters that govern the dynamics of the system are the Rayleigh ($Ra$), bulk Richardson ($Ri_b$), Prandtl ($Pr$), Stefan ($St$) and Rossby ($Ro$) numbers, and are defined as
\be
Ra = \frac{g \, \alpha \, \Delta T \, H^3}{\nu \, \kappa}{;} \quad Ri_b = \frac{g \, \alpha \, \Delta T}{F_p}{;} \quad Pr = \frac{\nu}{\kappa}{;}
\ee
\be
St = \frac{L_s}{C_p \, \Delta T}{;} \quad \mbox{and} \quad Ro = \sqrt{\frac{g \, \alpha \, \Delta T}{\Omega^2 \, H}},
\ee
where $C_p$ is the specific heat of the solid phase. These represent the ratios of buoyancy to viscous forces ($Ra$), of buoyancy to the {imposed 
horizontal body force per unit mass} ($Ri_b$),  of momentum to heat diffusivities ($Pr$),  of the latent to specific heats ($St$), and of rotational to convective time scales ($Ro$). Note that $St$ is defined based on the temperature difference in the liquid layer and not the solid layer. A bulk Reynolds number can be defined by combining $Ra$ and $Ri_b$, and is given by
\be
Re = \sqrt{\frac{Ra}{Ri_b  Pr}} \label{eq:bulk_Re}.
\ee
In cases where both rotation and shear are present, we define $\Sigma$ as the ratio of centrifugal to applied body forces,
\be
\Sigma = \frac{Ri_b}{Ro^2} = \frac{\Omega^2 H}{F_p}, \label{eq:Sigma}
\ee
such that mean shear dominates when $\Sigma < 1$.

The dimensionless temperature boundary conditions are:
\be
\theta(x, y, z=0,t) = 1 \quad \mbox{and} \quad \theta(x,y,z = H,t) = 0,
\ee
at the bottom and top surfaces, respectively, and
\be
\theta(x,y,z=h,t) = 0,
\ee
at the phase boundary. The no-slip and no-penetration boundary conditions for the velocity field remain unaltered after the non-dimensionalisation.

An important means of quantifying the response of the system is in terms of the Nusselt number, which is defined as the ratio of total heat flux to the heat flux due only to conduction. We define the Nusselt number as
\begin{equation}
Nu = {\sqrt{Ra Pr  St^2} \, h \, \frac{dh}{dt}},
\label{eqn:Nusselt}
\end{equation}
where {$h$ is the horizontally averaged height of the fluid layer.  It is the total heat flux scaled by the conductive heat flux over the liquid layer, whose depth is evolving.  This follows \cite{ravichandran2021}, although their Stefan number is the inverse of that defined here.  }

\section{Numerical method \label{sec:num-method}}

Equations \ref{eqn:mass_scaled} -- \ref{eqn:stefan_scaled}, along with the boundary conditions, are solved using the finite-volume solver \emph{Megha-5}, with {a volume penalization method} for the melting solid. The solver uses second-order central differences in space and a second-order Adams-Bashforth time-stepping scheme, and has been extensively validated for free-shear flows \citep{singhal2021reynolds, singhal2021virus} and Rayleigh-B\'enard convection \citep{ravichandran2020_cloud,ravichandran2020transient,ravichandran2021}. Details of the volume-penalization algorithm, including validation against the analytical solution for the purely conductive Stefan problem, tests of sensitivity to the penalization parameter, and the convergence of the solution under grid-refinement may be found in \citet{ravichandran2021}. {To test the results for grid independence in this study, simulations for the lowest $Ri_b$ cases were repeated with double the number of grid points in the vertical and the difference in the results was found to be negligible. Furthermore, we have verified that the body force prescribed in equation \ref{eqn:NS_liquid} produces the known analytical plane-Poiseuille velocity profile with an exponential relaxation timescale of $(\pi^2 \nu)^{-1}$ for the peak velocity.}

\section{Results}

Here, we present our results on the effects of mean shear, buoyancy and rotation on the evolution of the phase boundary. In order to highlight the effects of rotation, we first consider only the effects of mean shear and buoyancy, and later introduce rotation. Choosing $Pr > 1$ ensures that $Ro < 1$ without unduly increasing the resolution requirement for the Ekman boundary layers at the upper and lower surfaces. For this reason, we choose $Pr = 5$ for the cases where rotation is present \citep[see e.g.,][]{ravichandran2020transient}. In the absence of rotation, there is no qualitative difference in the melting dynamics for $Pr = 1$ and $5$. The choice of $Ra$ in this study reflects a balance between choosing a sufficiently large $Ra$ to observe the convective dynamics whilst minimizing the computational effort required to resolve the thermal boundary layers.

\begin{table*}
\begin{centering}
\begin{tabular}{c  c  c  c  c  c  c  c  c }
\hline
$Ro$ & $Pr$ & $Ri_b$ & $Re$ & $h_0$ & $(L_{x}\times L_{y})\times L_{z}$ & $N_{x},N_{y},N_{z}$ & $\Delta t$ &  Figures \tabularnewline
\hline
$\infty$ & $1$ & $\infty$ & $0$ & $1$ & $16^{2}\times2$ & $1024^{2}\times128$ & $7\times10^{-4}$ & \ref{fig:solid_liq_interface_E0}, \ref{fig:sum_chi_vs_t_E0} \tabularnewline
\hline 
$\infty$ & $1$ & $100$ & $35.4$ & $1$ & $16^{2}\times2$ & $1024^{2}\times128$ & $7\times10^{-4}$ & \ref{fig:solid_liq_interface_E0}, \ref{fig:sum_chi_vs_t_E0}\tabularnewline
\hline 
$\infty$ & $1$ & $10$ & $111.8$ &  $1$ & $16^{2}\times2$ & $1024^{2}\times128$ & $7\times10^{-4}$ & \ref{fig:solid_liq_interface_E0}, \ref{fig:Rep0d1_E0}, \ref{fig:hovmoller_E0}, \ref{fig:u_theta_vs_z_nonrot}, \ref{fig:sum_chi_vs_t_E0}  \tabularnewline
\hline 
$\infty$ & $1$ & $4$ & $176.8$ &  $1$ & $16^{2}\times2$ & $1024^{2}\times128$ & $7\times10^{-4}$ & \ref{fig:solid_liq_interface_E0}, \ref{fig:Rep0d1_E0}, \ref{fig:hovmoller_E0}, \ref{fig:u_theta_vs_z_nonrot},  \ref{fig:sum_chi_vs_t_E0}  \tabularnewline
\hline 
$\infty$ & $1$ & $1$ & $353.6$ &  $1$ & $16^{2}\times2$ & $1024^{2}\times256$ & $2.8\times10^{-4}$ & \ref{fig:solid_liq_interface_E0}, \ref{fig:Rep1_E0}, \ref{fig:u_theta_vs_z_nonrot}, \ref{fig:sum_chi_vs_t_E0} \tabularnewline
\hline
\hline
$Ro$ & $Pr$ & $Ri_b$ & $\Sigma$ & $h_0$ & $(L_{x}\times L_{y})\times L_{z}$ & $N_{x},N_{y},N_{z}$ & $\Delta t$ &  Figures \tabularnewline
\hline
$0.6325$ & $5$ & $100$ & $250$ &  $1$ & $16^{2}\times2$ & $1024^{2}\times128$ & $7\times10^{-4}$ & \ref{fig:solid_liq_interface_E5em4}, \ref{fig:sum_chi_vs_t_E5em4} \tabularnewline
\hline
$0.6325$ & $5$ & $10$ & $25$ &  $1$ & $16^{2}\times2$ & $1024^{2}\times128$ & $7\times10^{-4}$ & \ref{fig:solid_liq_interface_E5em4}, \ref{fig:sum_chi_vs_t_E5em4} \tabularnewline
\hline 
$0.6325$ & $5$ & $1$ & $2.5$ &  $1$ & $16^{2}\times2$ & $1024^{2}\times128$ & $7\times10^{-4}$ &  \ref{fig:solid_liq_interface_E5em4}, \ref{fig:Rep1_E5em4}, \ref{fig:sum_chi_vs_t_E5em4} \tabularnewline
\hline 
$0.6325$ & $5$ & $0.5$ & $1.25$ &  $1$ & $16^{2}\times2$ & $1024^{2}\times128$ & $7\times10^{-4}$ & \ref{fig:solid_liq_interface_E5em4}, \ref{fig:sum_chi_vs_t_E5em4}\tabularnewline
\hline 
$0.6325$ & $5$ & $0.2$ & $0.5$ &  $0.5$ & $8^{2}\times2$ & $512^{2}\times128$ & $7\times10^{-4}$  &  \ref{fig:solid_liq_interface_Rep5_10_h0d25},  \ref{fig:uv_mean_Rep10}, \ref{fig:sum_chi_vs_t_E5em4_h0d25}\tabularnewline
\hline 
$0.6325$ & $5$ & $0.1$ & $0.25$ &  $0.5$ & $8^{2}\times2$ & $512^{2}\times128$ & $7\times10^{-4}$ & \ref{fig:wT_Rep10_h0d25}, \ref{fig:solid_liq_interface_Rep5_10_h0d25},  \ref{fig:uv_mean_Rep10}, \ref{fig:hovmoller_E5em4_h0d25}, \ref{fig:sum_chi_vs_t_E5em4_h0d25}, \ref{fig:u_theta_vs_z_rot_h0d25}\tabularnewline
\hline 
\end{tabular}
\par\end{centering}
\caption{\label{tab:cases} List of cases reported here. In all cases the Rayleigh number is $Ra=1.25\times10^5$, \tb{the Stefan number $St=1$} and the penalisation parameter is $\eta=1.4\times10^{-3}$. \tb{We also give the bulk Reynolds number $Re$ for the non-rotating cases (equation \ref{eq:bulk_Re}) and the parameter $\Sigma$ for the rotating cases (equation \ref{eq:Sigma}).} The solutions obtained are independent of the grid resolution, time step $\Delta t$ and volume penalisation parameter $\eta$. }
\end{table*}

\subsection{Zero rotation ($Ro=\infty$) \label{subsec:Zero-rotation}}

When rotational effects are absent, the evolution of the phase boundary is affected by the combined action of the mean shear flow and convection. The relative strength of {mean shear to buoyancy} is quantified using $Ri_b$: buoyancy dominates when $Ri_b \gg 1$, and mean shear dominates when $Ri_b \ll 1$. We explore the range $Ri_b \in \left[1, \infty\right)$ in this section, with the other dimensionless parameters fixed at $Ra=1.25\times10^5$ and $Pr = St = 1$.

\begin{figure}
\begin{centering}
\includegraphics[width=0.49\columnwidth]{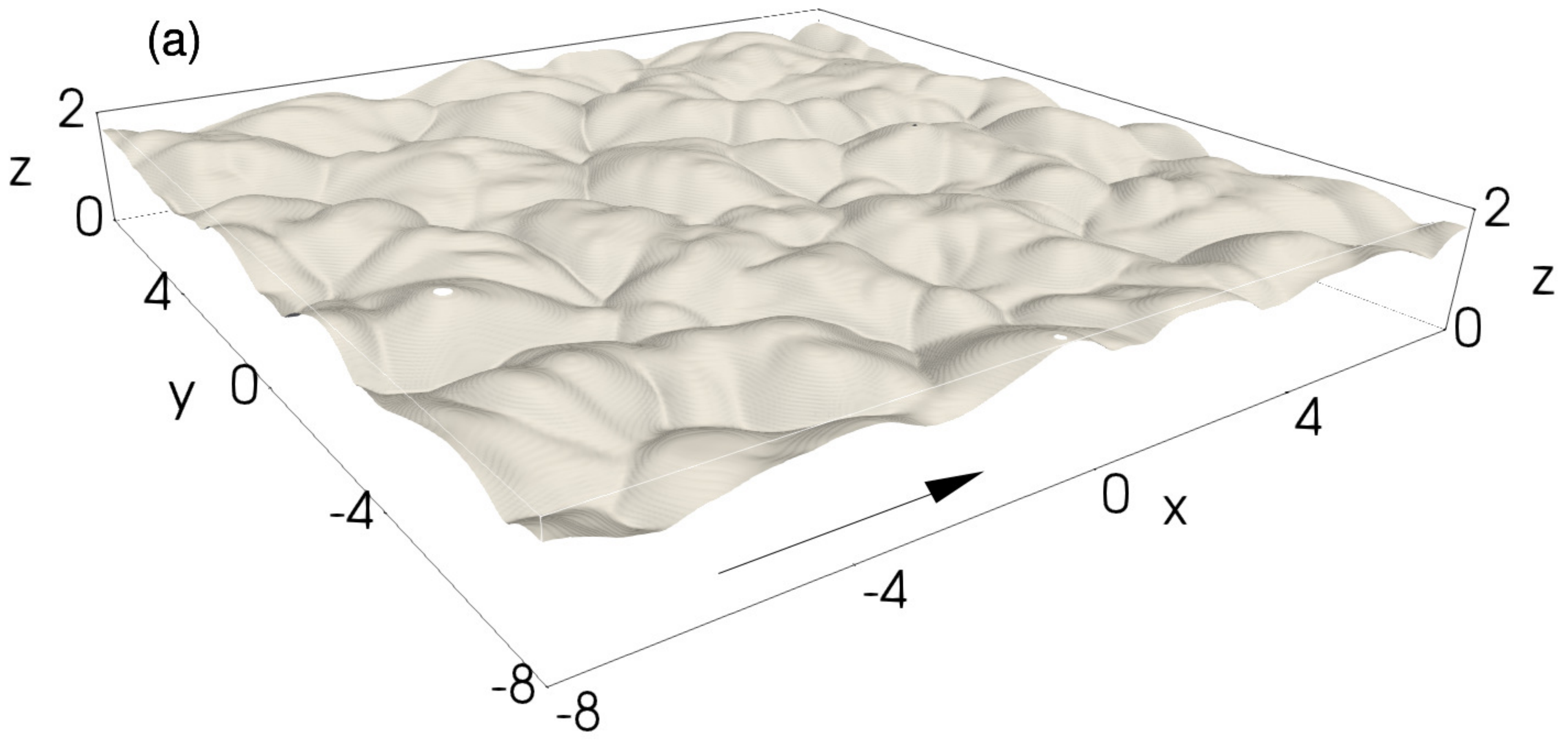}
\includegraphics[width=0.49\columnwidth]{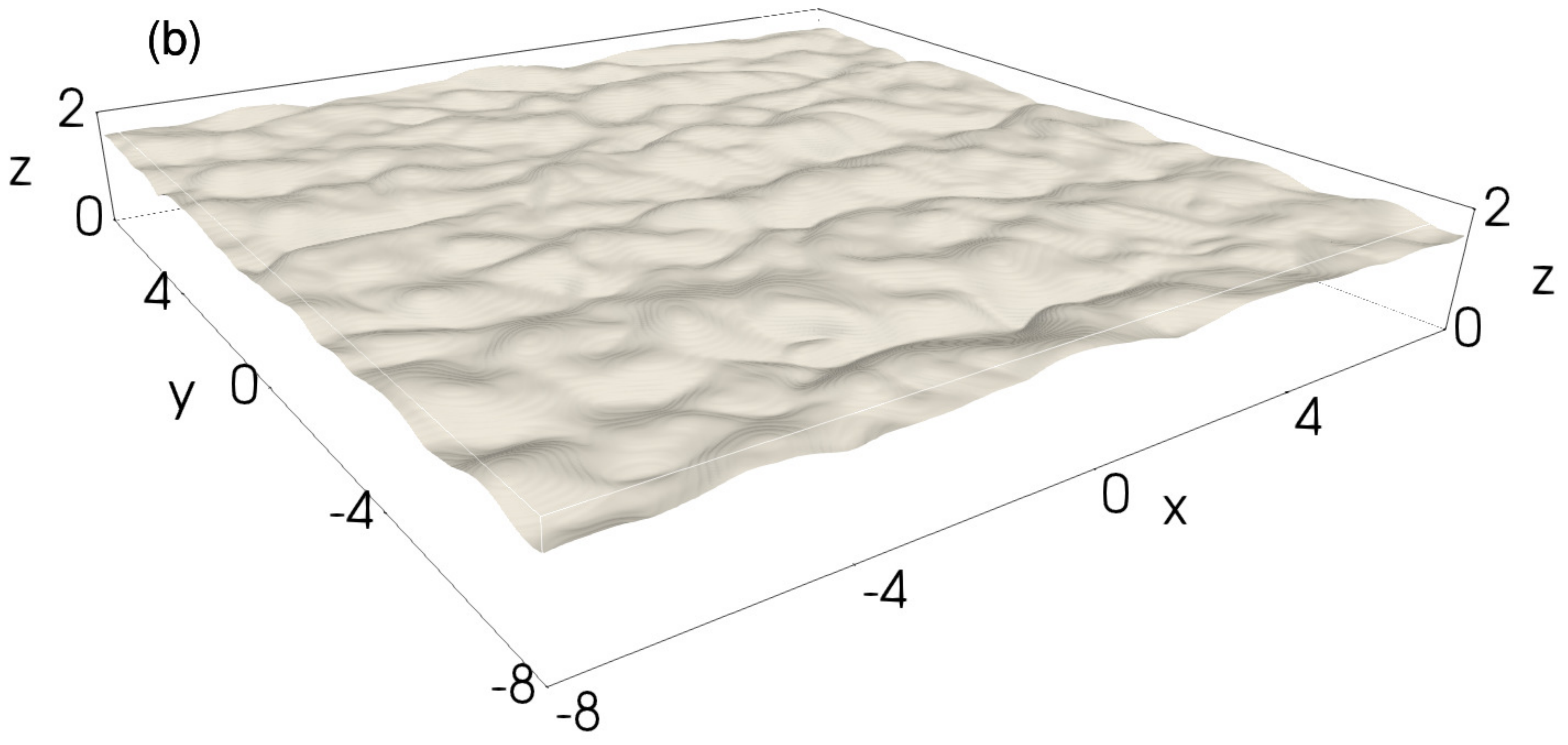}
\par\end{centering}
\begin{centering}
\includegraphics[width=0.49\columnwidth]{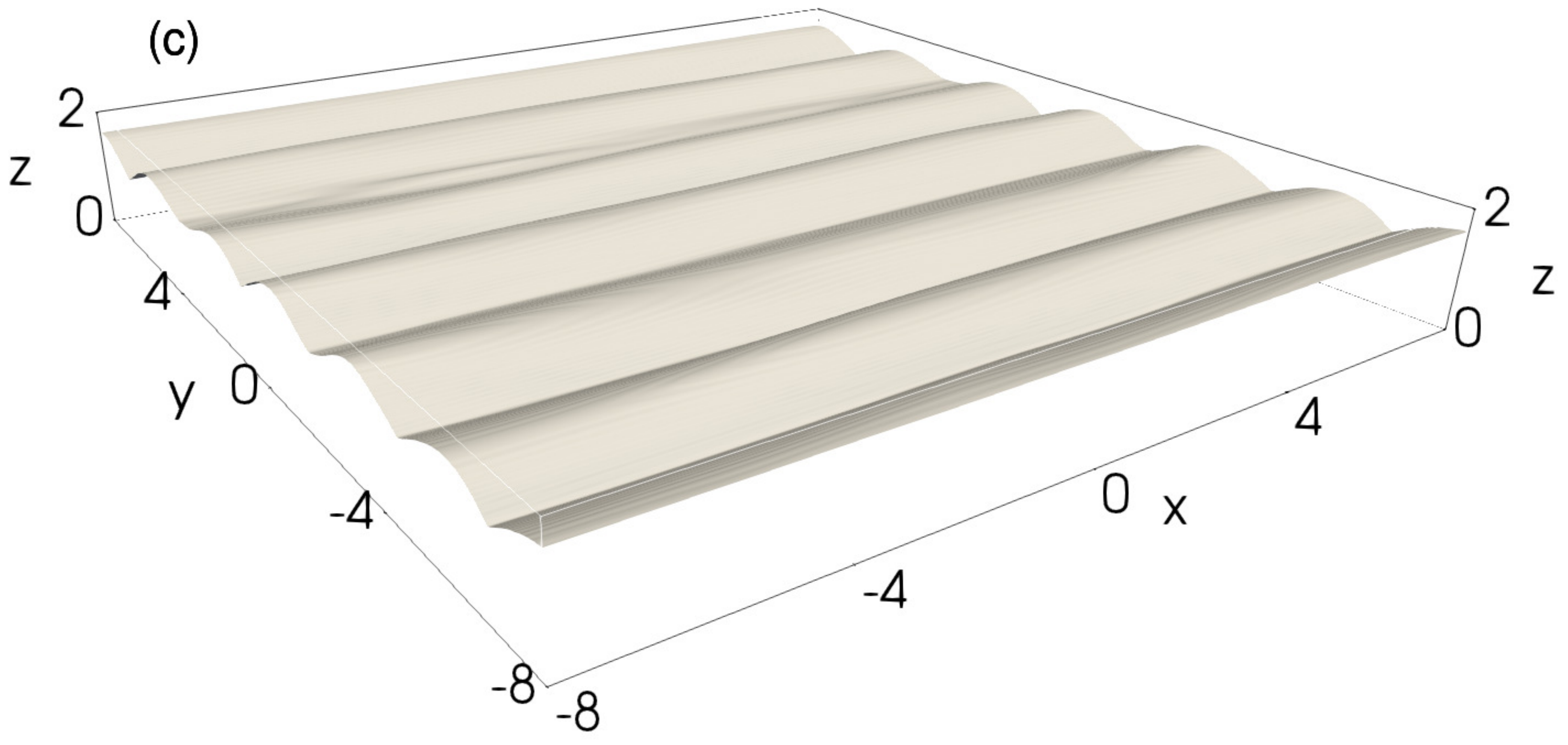}
\includegraphics[width=0.49\columnwidth]{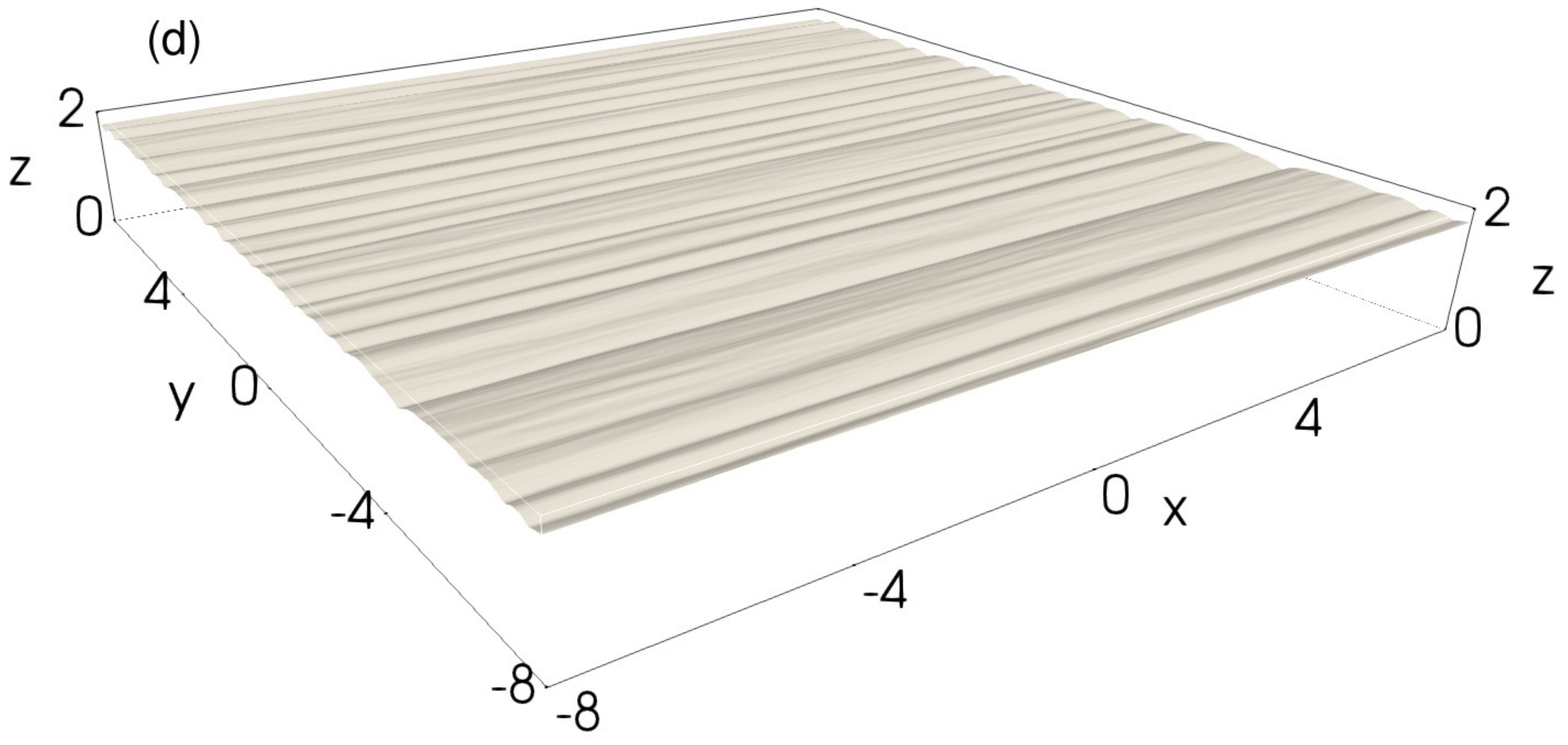}
\par\end{centering}
\caption{\label{fig:solid_liq_interface_E0} The solid liquid interface viewed from the solid side for $Ro=\infty$, $Ra=1.25\times10^5$, $Pr=1$, and $St=1$ at $t=85$ for (a) $Ri_b = \infty$,  (b) $Ri_b=100$, (c) $Ri_{b}=10$, (d) $Ri_{b}=1$. The arrow in (a) denotes the direction of the shear flow along the $x$ direction.
For $Ri_b=\infty$, the melting creates dome-like voids in the solid, as seen in (a). Nascent streamwise alignment of the melting pattern appears for $Ri_{b} = 100$ in (b). For smaller $Ri_b$, the formation of quasi-two dimensional grooves separated by sharp furrows is seen in (c) and (d).  {The mean liquid layer depths are comparable in (a-c), while the liquid depth in (d) is about $20\%$ larger (see figure \ref{fig:sum_chi_vs_t_E0}).}} 
\end{figure}

In figures \ref{fig:solid_liq_interface_E0}(a) -- \ref{fig:solid_liq_interface_E0}(d), we show the effects of increasing {the} strength of {the} mean shear flow on the phase boundary at $t=85$.

When mean shear is absent ($Ri_b = \infty$), convective motions tend to create dome-like features at the boundary, which `lock-in' the convection cells. This is seen in figure \ref{fig:solid_liq_interface_E0}(a). Such features have been investigated in detail in both two \citep{favier2019, purseed2020} and three \citep{esfahani2018} dimensions. As the strength of the mean shear is increased (decreasing $Ri_b$), quasi two-dimensional {corrugations with a wavelength transverse to the direction of the shear flow} begin to emerge at the phase boundary, as seen in figures \ref{fig:solid_liq_interface_E0}(b) -- \ref{fig:solid_liq_interface_E0}(d). The shear flow, which is {parallel to} the $x$-axis, tends to homogenize the phase boundary along {the flow} direction. However, it does not affect the corrugations perpendicular to {the flow; parallel to the $y$-axis} \citep{TW2019}. Such features have also been observed in the recent study of \citet{couston2020}. The effect of the mean shear flow here is consistent with that in two dimensions, {where the corrugations of a phase boundary decrease in amplitude as a result of shear} \citep{toppaladoddi2020}. 

{The quasi two-dimensional features at the phase boundary are due to the emergence of the longitudinal rolls, which are the preferred form of convection when $Ra > Ra_c$ and the $Re$ is not too large \citep{clever1991}. These longitudinal rolls can be discerned in the contours of the vertical component of the velocity and temperature. While these effects begin to appear for $Ri_b=100$ (not shown), they are clearly evident for $Ri_b=10$, as shown in figure \ref{fig:Rep0d1_E0}.  
\begin{figure}
\includegraphics[width=1\columnwidth]{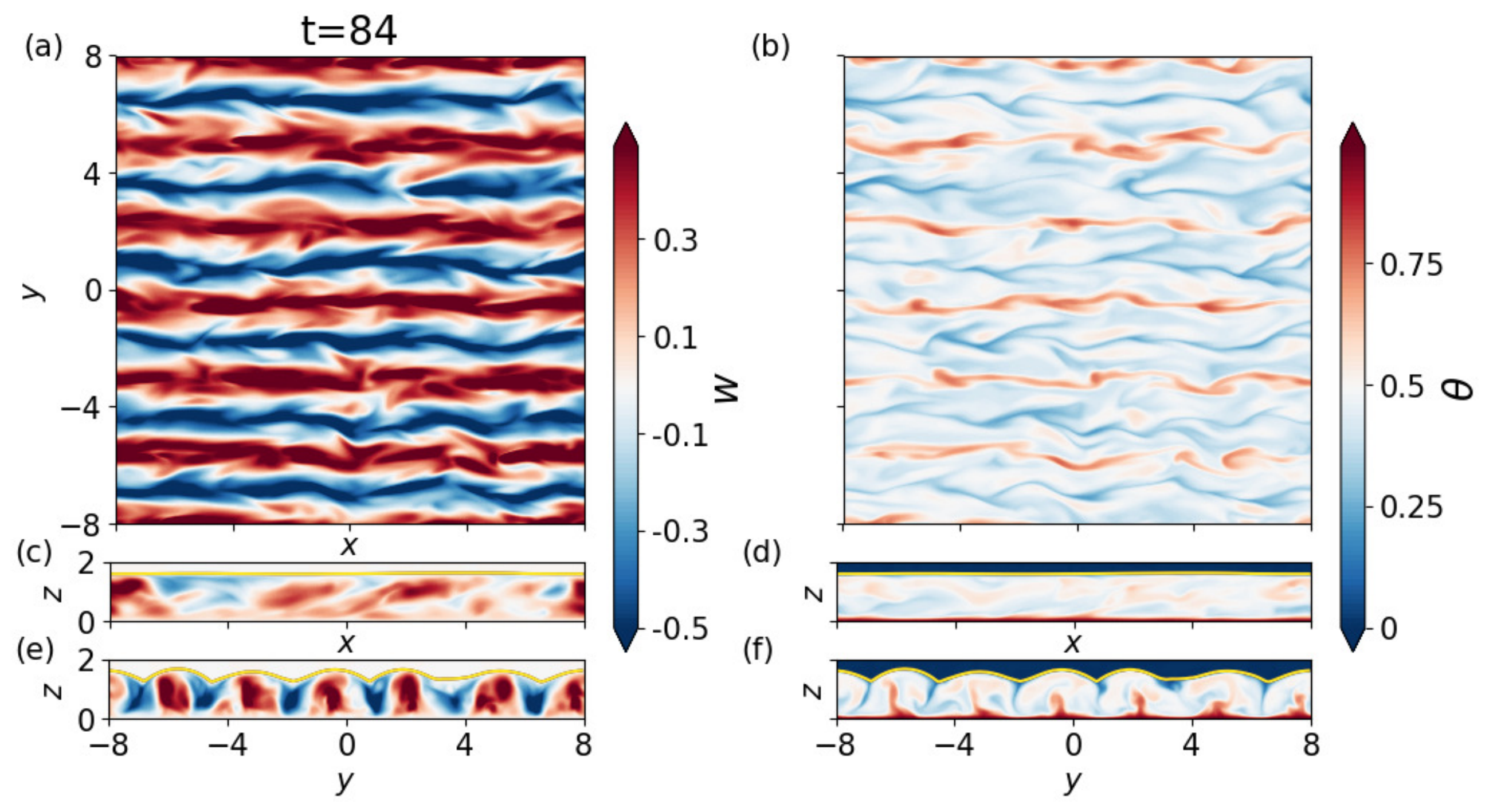}
\caption{\label{fig:Rep0d1_E0} Contours of vertical velocity $w$ and temperature $\theta$ at the horizontal section $z=H/2$ (a,b) and the vertical sections $y=0$ (c,d) and $x=0$ (e,f) at t=84. The solid-liquid interface is shown in the vertical sections as a solid yellow line. The flow parameters are $Ri_{b}=10$, $Ro=\infty$, $Ra=1.25\times10^5$, $Pr=1$, and $St=1$. Here, quasi two-dimensional corrugations with a wavelength transverse to the direction of the shear flow are clearly associated with longitudinal roll structures in the flow.}
\end{figure}
Vertical sections of the flow in figures \ref{fig:Rep0d1_E0}(c,d)  show that the interface is homogeneous in the direction of the shear flow}. {These phase boundary patterns are qualitatively different from those observed experimentally by \cite{Gilpin1980}.
This is due to the fact that in their experiments the $Re$ based on the boundary-layer thickness was $O(10^4)$, whereas in our simulations the bulk $Re$ is $O(10^2-10^3)$, and the wavelength of transverse patterns is $O(10^3)$ times the viscous sublayer thickness \citep{Claudin2017} and hence longer than our simulation domain. Longitudinal features are also observed in flows over eroding \citep{Allen1971} and dissolving boundaries \citep{cohen2016, cohen2020}, where the associated grooves eventually merge to produce scallops \citep[see also][]{bushuk2019}.} Another interesting feature to note is that for {$Ri_b = 10$}, the corrugations at the phase boundary become more regular in the $y-z$ plane, which is shown in figure \ref{fig:Rep0d1_E0}(f). This shows that the mean shear flow inhibits vertical motions, which is similar to its effects in two dimensions \citep{toppaladoddi2020}. 

\begin{figure}
\includegraphics[width=0.5\columnwidth]{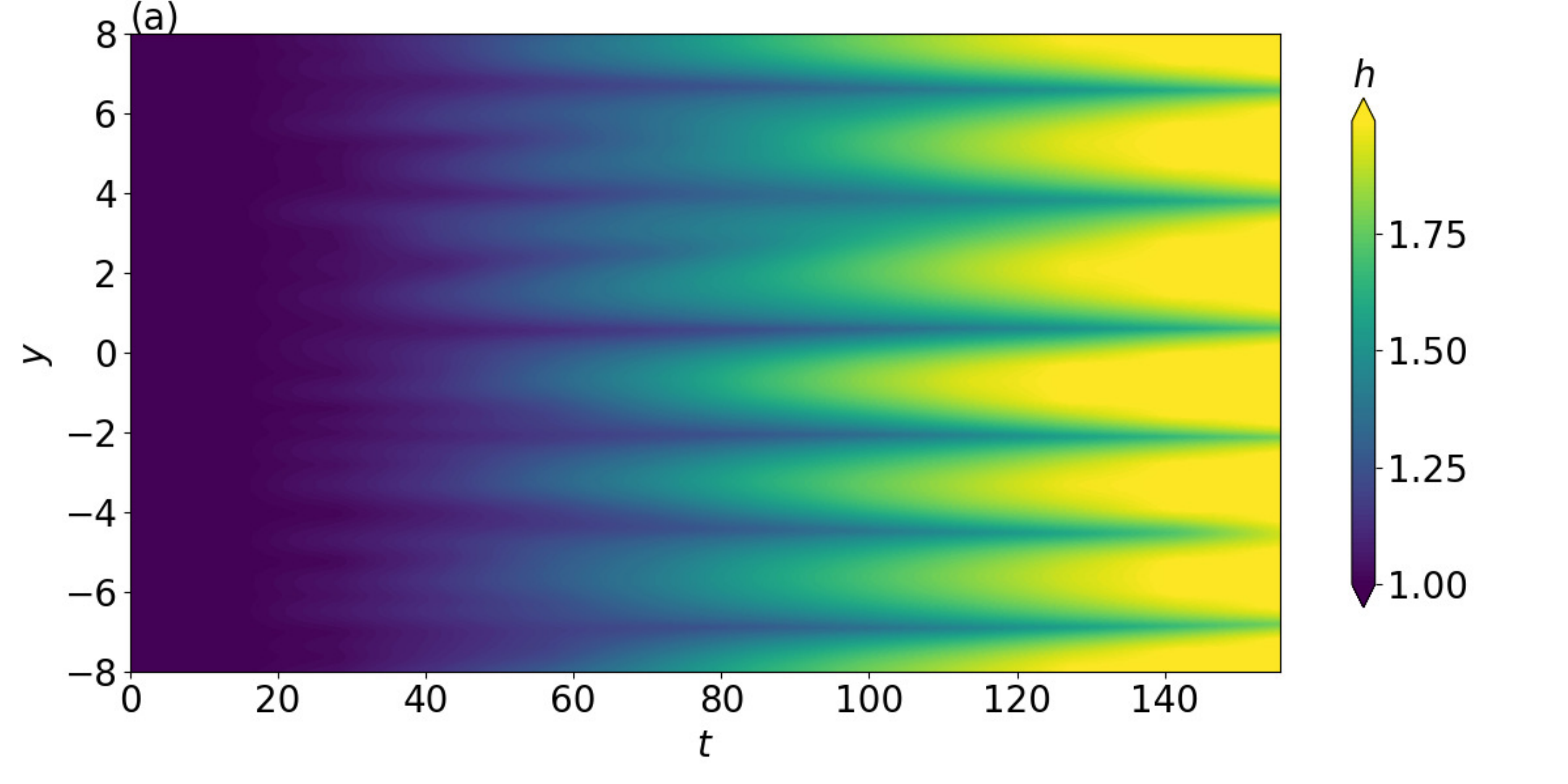}
\includegraphics[width=0.5\columnwidth]{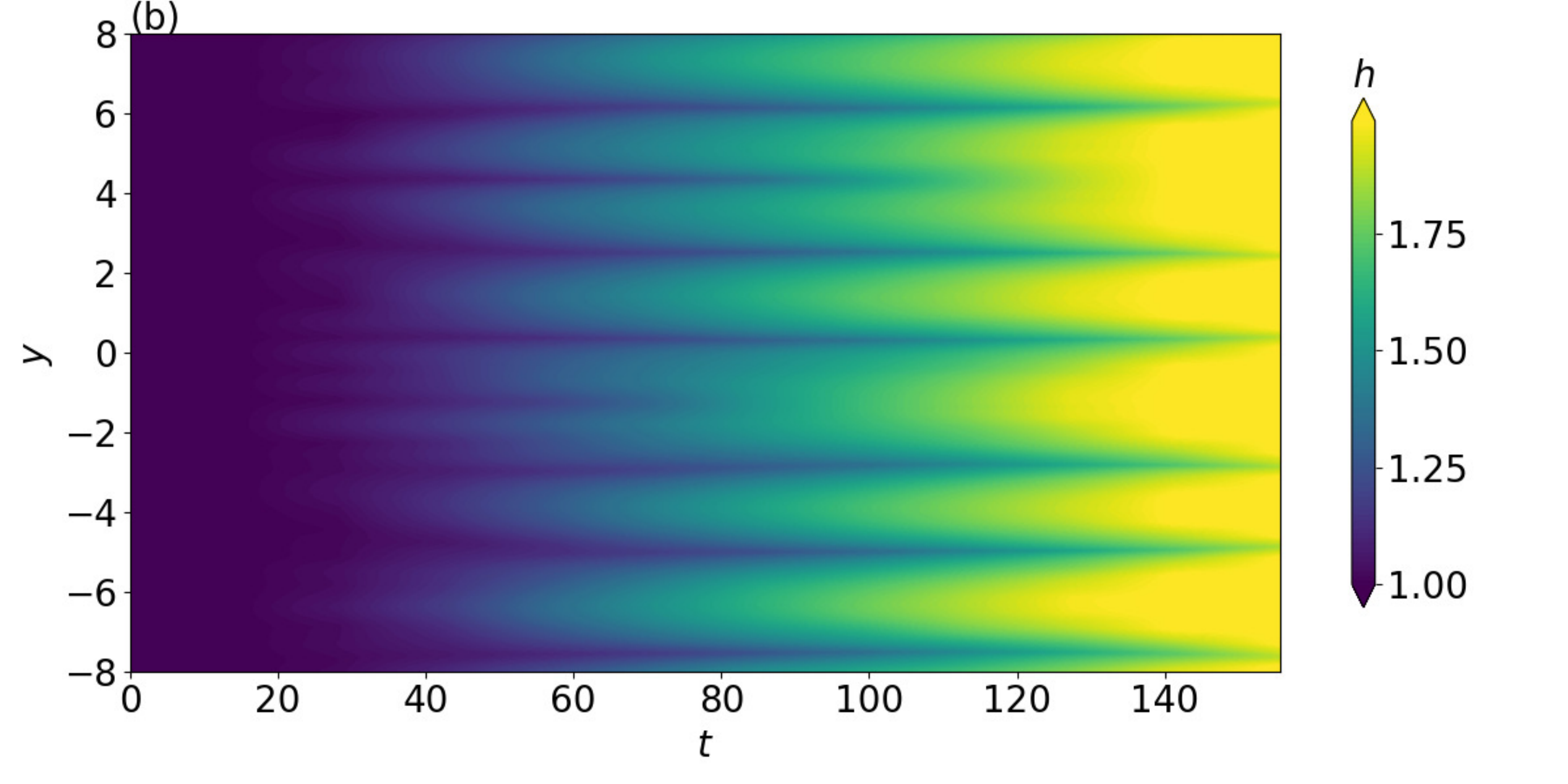}
\caption{\label{fig:hovmoller_E0} {\Hovmoller plots of the fluid height at $x=0$, plotted as a function of $(t,y)$ for (a) $Ri_b=10$ showing the emergence of $6$ domes of nearly equal widths (of about $2.66$), and (b) $Ri_b=4$ showing the emergence of $3$ domes each of two distinct sizes. In both instances, the domes maintain their horizontal positions.}}
\end{figure}

{The longitudinal rolls persist down to $Ri_b \approx 4$, below which they start to merge. The effects of the convective rolls on the melting morphology are shown in the \Hovmoller diagrams of the liquid layer depth as a function of $y$ and $t$ in figure \ref{fig:hovmoller_E0}.  At $t=150$,
for $Ri_b = 10$ the six corrugations have a wavelength of approximately 2.66, which is slightly larger than twice the initial fluid depth, whereas for $Ri_b=4$ the merging and coarsening of corrugations is evident (cf. figures \ref{fig:hovmoller_E0}(a) and (b)).  Prior to this time, the dynamics of the corrugations depends on $Ri_b$, highlighting the importance of the specific history in controlling the interfacial structure.  The coupling of the flow structures and the melting morphology is thus seen more starkly here than in convection without shear in non-rotating \citep[e.g.][]{esfahani2018} and rotating cases \citep{ravichandran2021}, where there are there domes, like those in figure \ref{fig:solid_liq_interface_E0}(a), rather than longitudinal structures, whose average size increases continuously as the solid melts.}

\begin{figure}
\includegraphics[width=1\columnwidth]{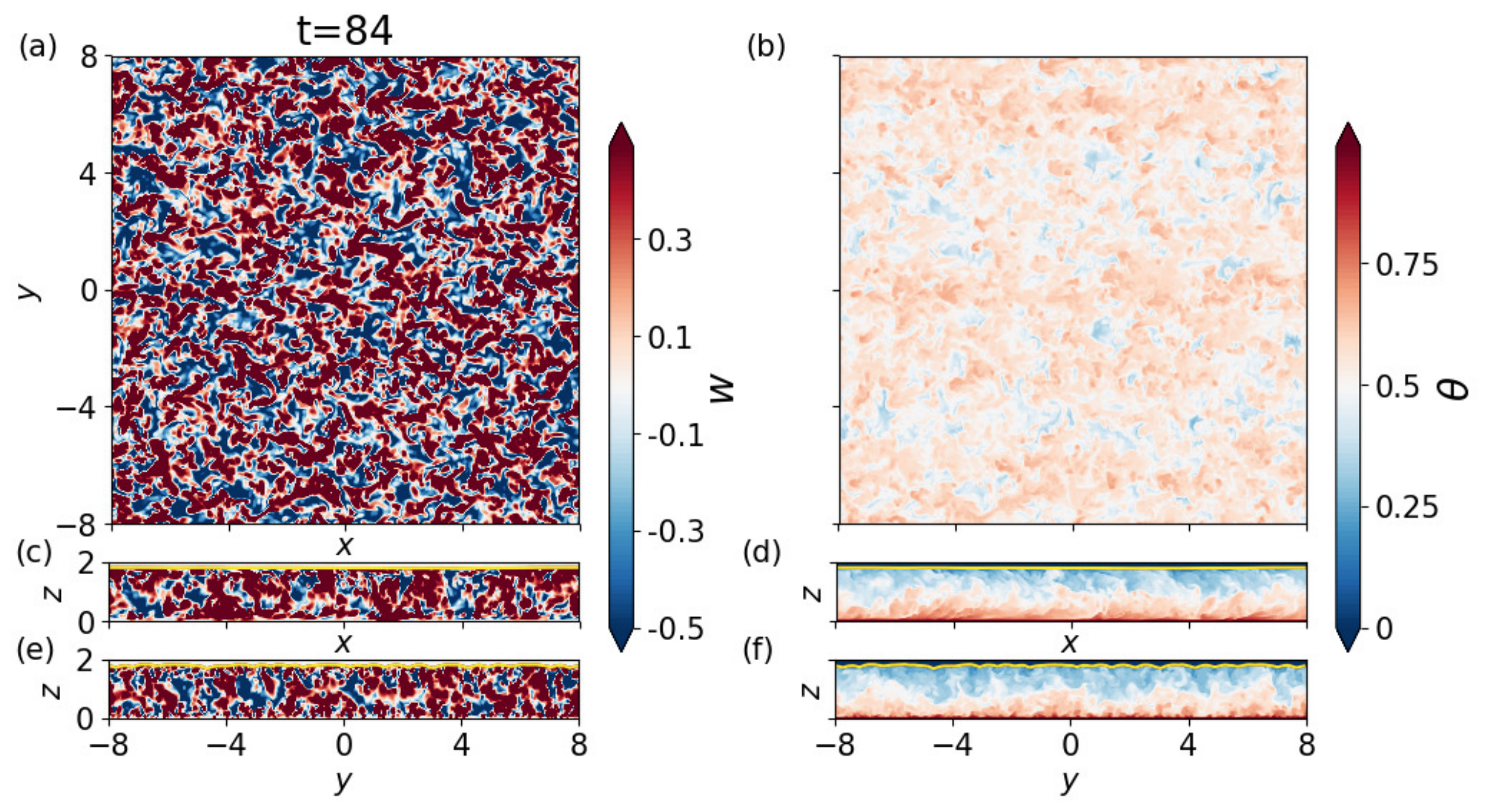}
\caption{\label{fig:Rep1_E0} {Contours of vertical velocity $w$ and temperature $\theta$ as in figure \ref{fig:Rep0d1_E0} and for the same flow parameters except $Ri_{b}=1$}. The increased shear, compared to $Ri_b=10$, causes the flow to become fully three-dimensional. This greatly increases the heat transfer rate and the Nusselt number (see also figure \ref{fig:sum_chi_vs_t_E0}). As a result, the grooves are \emph{less prominent} ({c.f.} figure \ref{fig:solid_liq_interface_E0}).}
\end{figure}

\begin{figure}
\begin{centering}
\includegraphics[width=1\columnwidth]{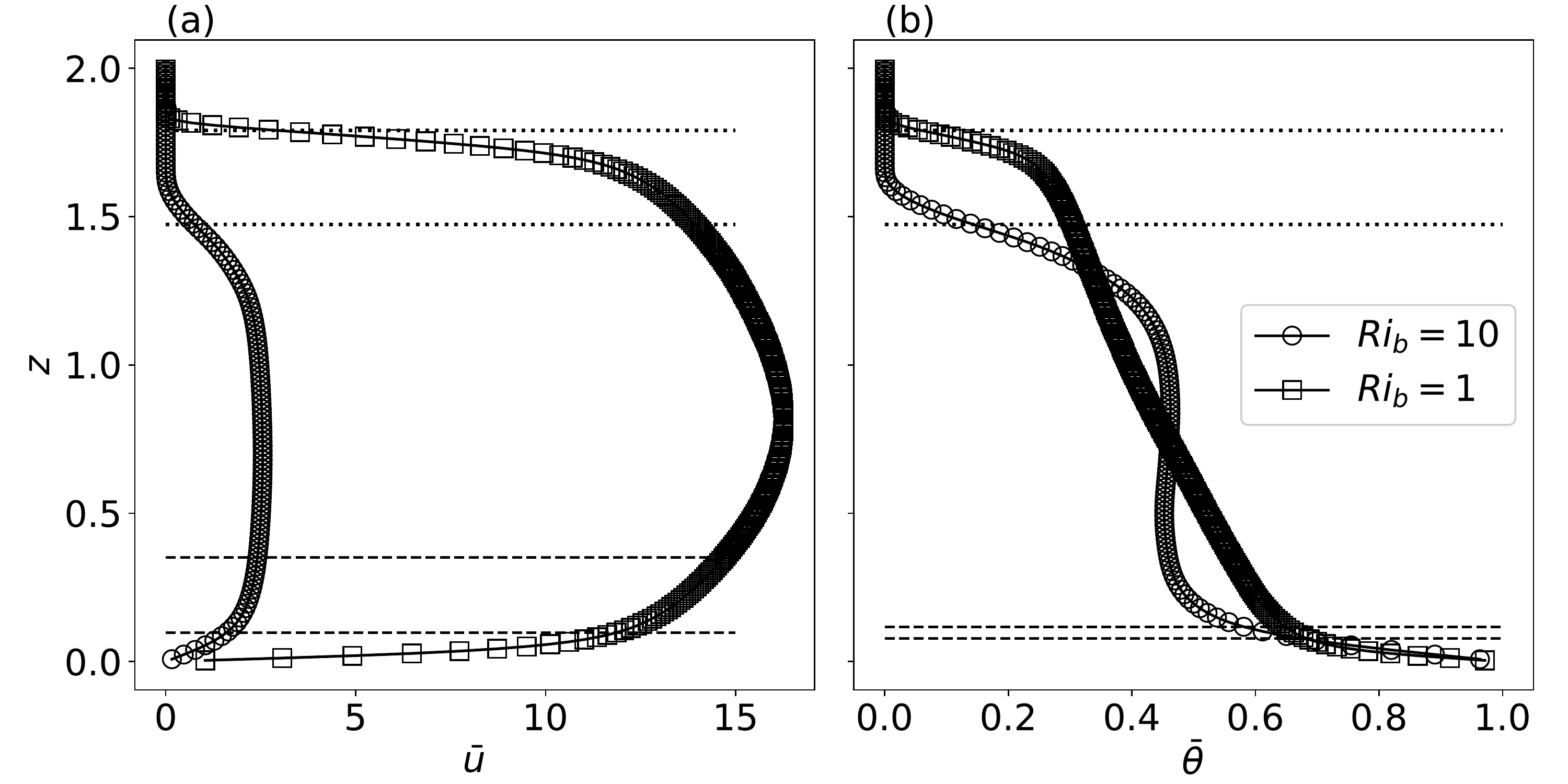}
\includegraphics[width=0.6\columnwidth]{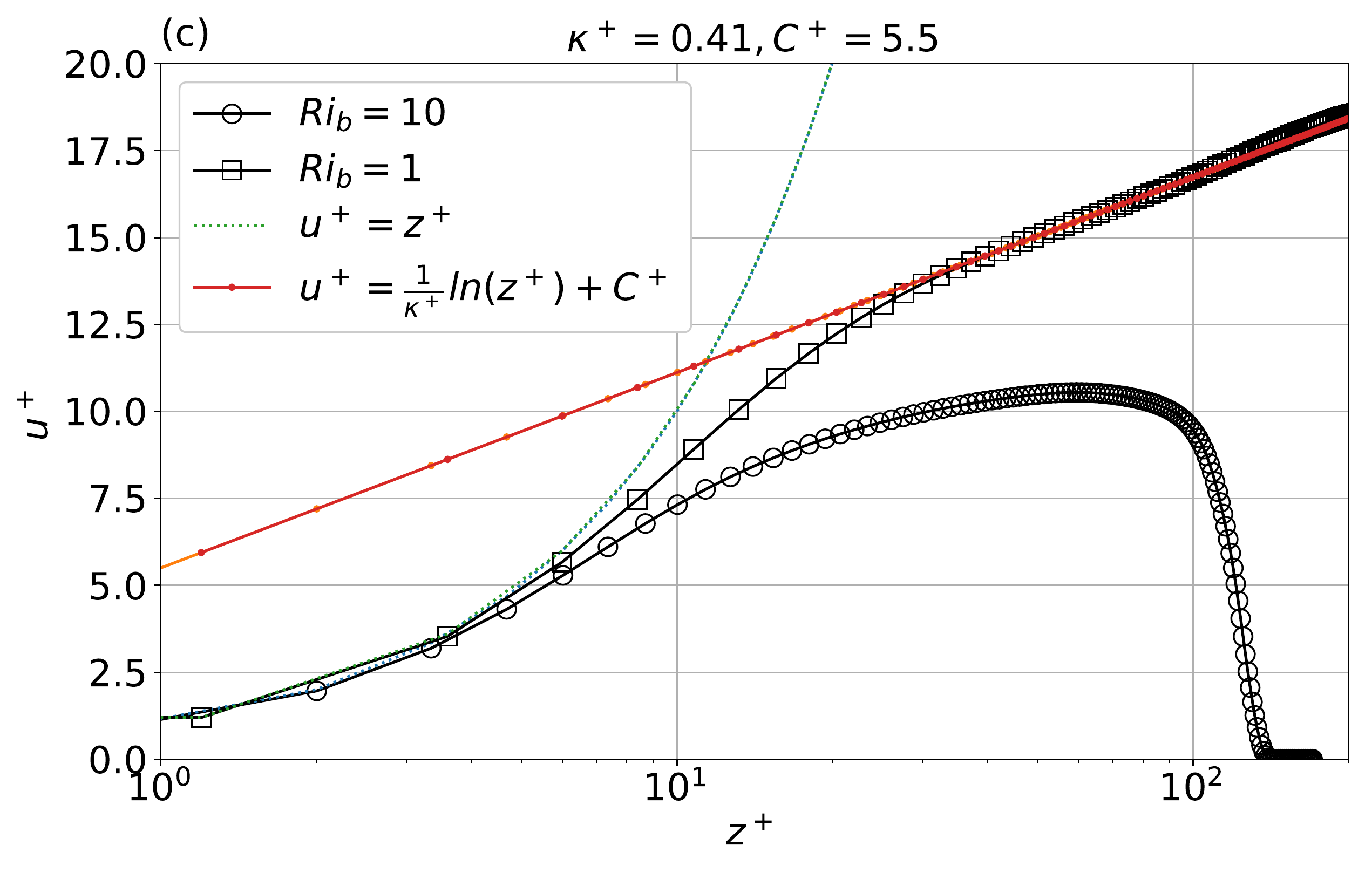}
\par\end{centering}
\caption{\label{fig:u_theta_vs_z_nonrot} Profiles of the horizontally averaged (a) horizontal velocity $\bar{u}$ and (b) temperature $\bar{\theta}$ at $t\approx85$ for $Ri_b=1$ and $Ri_b=10$ , with $Ro=\infty$, $Ra=1.25\times10^5$, $Pr=1$, $St=1$. The velocity and thermal boundary layer thicknesses ($30$ wall units for the velocity) are shown using dashed lines; \tb{the average height of the fluid layer is shown using dotted lines;} the boundary layers are thinner for smaller $Ri_b$. For sufficiently small $Ri_b$, a turbulent boundary layer develops at the lower (smooth) boundary, with the well-known logarithmic velocity profile for $z^+ \gtrsim 30$, \tb{where $z^+ = z/\delta_\tau$ is the vertical coordinate scaled in terms of the wall-friction thickness $\delta_\tau = \nu / u_\tau$, and $u_\tau = \left[\nu \left(\partial \bar{u} / \partial z \right)_{z=0} \right]^{1/2}$ is the friction velocity.} As a result, for large $Ri_b$, convection dominates, and the fluid bulk is nearly isothermal; for small $Ri_b$, on the other hand, the fluid bulk has a constant temperature gradient. The friction Reynolds numbers \tb{$Re_\tau = u_\tau H / \nu$}, based on the initial liquid height, are $85.4$ and $306.5$ for $Ri_b=10$ and $Ri_b=1$, respectively.}
\end{figure}

When $Ri_b$ is decreased further, the longitudinal rolls lose coherence and the bulk flow becomes turbulent and fully three-dimensional. This is shown in figures \ref{fig:Rep1_E0} and \ref{fig:u_theta_vs_z_nonrot}(a) and (b) for $Ri_b = 1$.  Nonetheless, remnants of the longitudinal rolls can still be seen along the phase boundary parallel to the $y$-axis (figures \ref{fig:Rep1_E0}(e) and (f)), which is homogeneous along the $x$ (figures \ref{fig:Rep1_E0}(c).
Clearly, as seen in figure \ref{fig:u_theta_vs_z_nonrot}(c), the classical log layer of a turbulent shear flow is responsible for a significant temperature gradient in the bulk.  
In contrast, for $Ri_b=10$, the bulk is nearly isothermal as expected in convection-dominated flows.

\begin{figure}
\begin{centering}
\includegraphics[width=1\columnwidth]{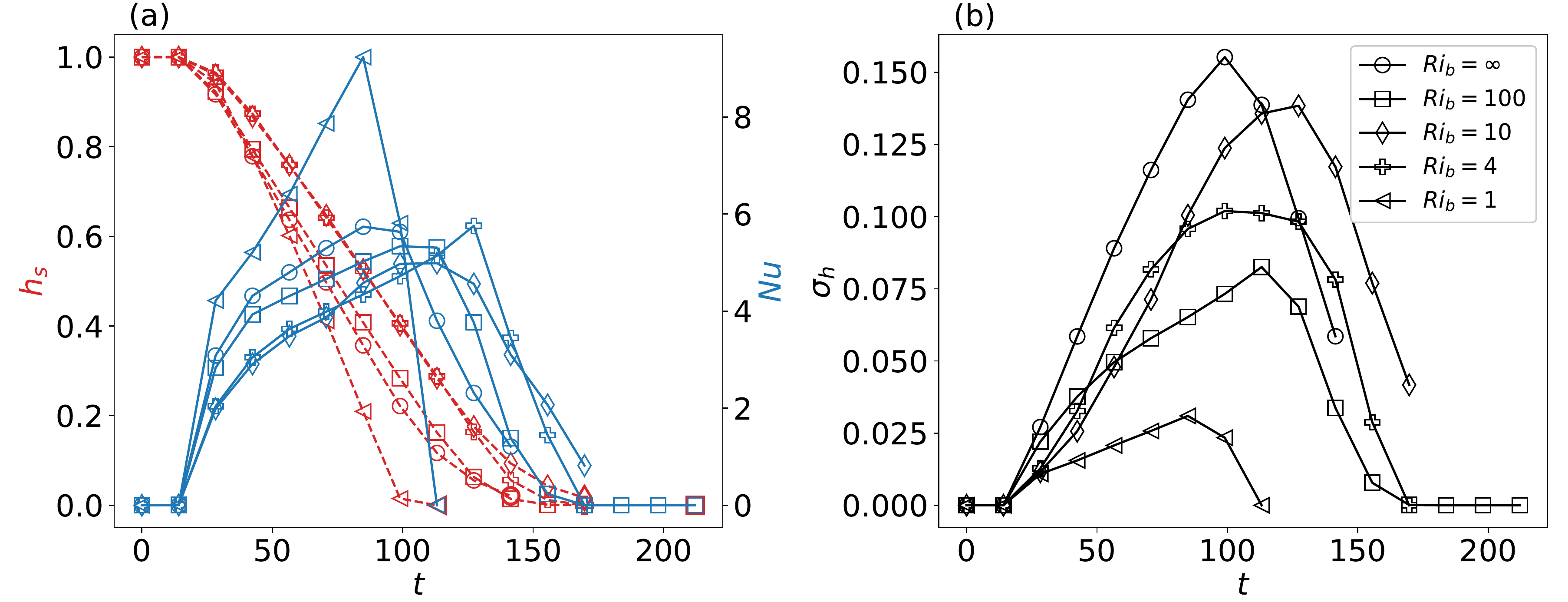}
\par\end{centering}
\caption{\label{fig:sum_chi_vs_t_E0}{ (a) The average thickness of the solid layer $h_s$ (red curves, left axis) and the melting Nusselt number (from eq. \ref{eqn:Nusselt}; blue curves, right axis); and (b) the roughness of the solid-liquid interface $\sigma_h$ for $Ro=\infty$, $Ra=1.25\times10^5$, $Pr=1$, {$St=1$}, and varying $Ri_b$. The rate of melting, and thus the Nusselt number, varies non-monotonically with $Ri_b$, as seen in (a). Similarly, the interface roughness also varies non-monotonically with $Ri_b$. (See also figure \ref{fig:max_Nu_stdh_vs_Rib}.)}}
\end{figure}

\begin{figure}
\begin{centering}
\includegraphics[width=1\columnwidth]{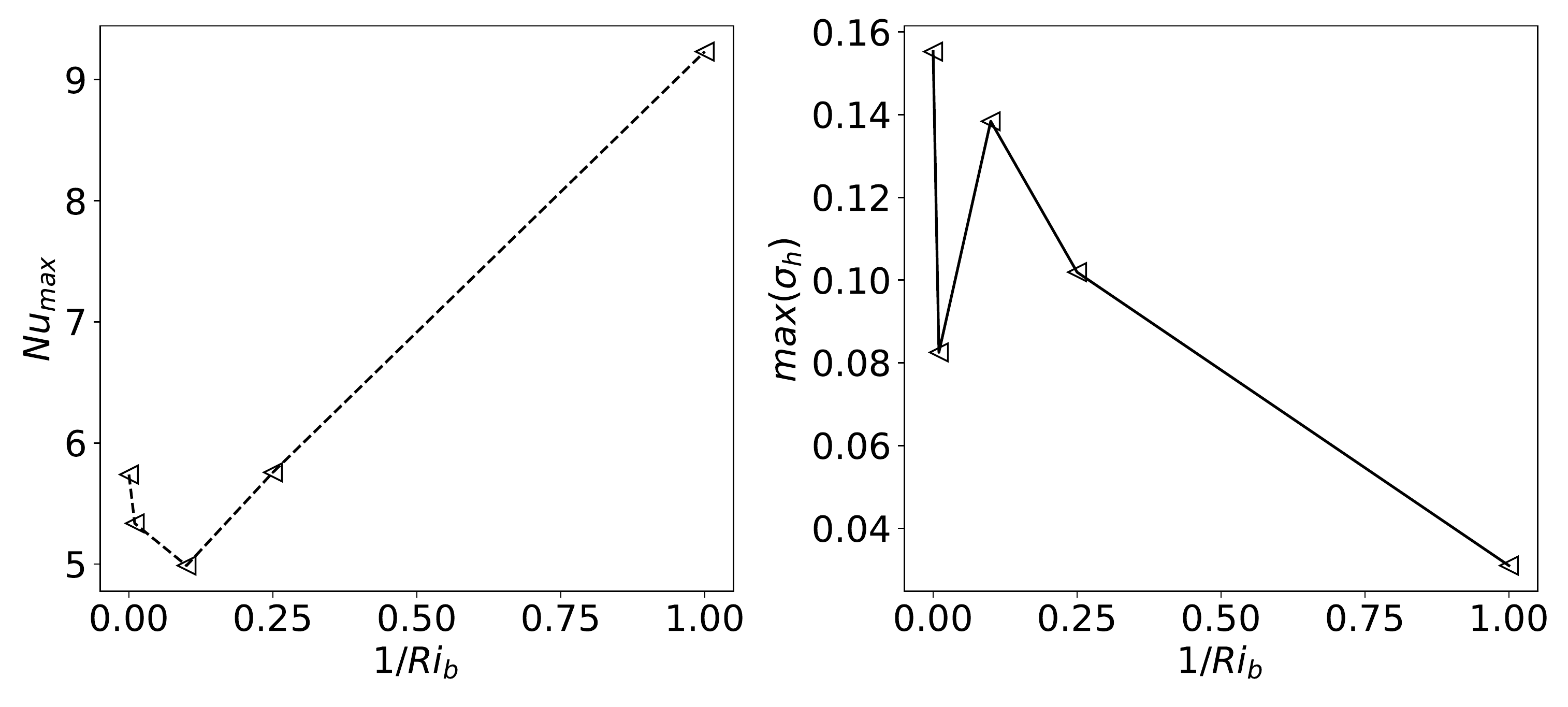}
\par\end{centering}
\caption{\label{fig:max_Nu_stdh_vs_Rib} The maximum values of (a) $Nu$ and (b) $\sigma_h$ in figure \ref{fig:sum_chi_vs_t_E0} as functions of $Ri_b$.}
\end{figure}

{The melting rates and interface roughness are functions of the flow properties and, as we have seen from figure \ref{fig:hovmoller_E0}, can also determine the behaviour of flow structures.} In figure \ref{fig:sum_chi_vs_t_E0}(a) we show the time evolution of $h_s$, and $Nu$ for different $Ri_b$. 
The melting rate of the solid, ascertained from the slopes of the $h_s(t)$ curves, is a non-monotonic function of $Ri_b$. When $Ri_b$ is reduced from $\infty$ to $4$, the rate of melting decreases. However, with a further decrease in $Ri_b$ {from $4$ to $1$}, there is an increase in the melting rate. This non-monotonicity results from the relative dominance of shear flow and convection in the dynamics of the system, \tb{and can also be seen in figures \ref{fig:sum_chi_vs_t_E0} and \ref{fig:max_Nu_stdh_vs_Rib}.}
For $Ri_b = 100$, the mean shear flow is relatively weak, but still acts to inhibit vertical motions. This results in decreased heat transport relative to the $Ri_b = \infty$ case. This trend continues for $Ri_b = 10$, where the vertical motions are further inhibited. However, a further decrease in $Ri_b$ to $1$ results in the mean shear flow becoming dominant, and the $Nu$ and melting rate increase due to the action of forced convection. These effects of decreasing $Ri_b$ can be seen by comparing figures \ref{fig:Rep0d1_E0}(f) and \ref{fig:Rep1_E0}(f), where the flow field transitions from having distinct plume structures to a more chaotic 3-D flow. The effects are qualitatively consistent with those observed in the study of pure Rayleigh-B\'enard-Poiseuille flow in 3-D {by \cite{toschi2014}}.

The roughness of the solid-liquid interface, defined here as the standard deviation $\sigma_h$ of the fluid height $h(x,y)$, is plotted in figure \ref{fig:sum_chi_vs_t_E0}(b). Similar to the melting rate, we see that the roughness also varies non-monotonically with increasing $Ri_b$, and is largest in the absence of shear, $Ri_b=\infty$. For $Ri_b = 100$, the incipient homogenisation of the flow structures, and thus the melting morphology, causes a decrease in $\sigma_h$. The emergence of the streamwise corrugations commensurate with convective rolls offsets the decrease of the roughness. As a result, an increase in the maximum of $\sigma_h$ is seen at $Ri_b=10$. For $Ri_b\leq10$, the dominance of shear and the resulting turbulent flow lead to less prominent morphological features and hence a reduction in $\sigma_h$. Furthermore, we note that, by comparing figures \ref{fig:sum_chi_vs_t_E0}(a) and (b), the roughness and the Nusselt number reach their maximal values as the solid-liquid interface reaches the upper boundary, as seen in the absence of shear \citep{ravichandran2021}.

\subsection{The effects of rotation} \label{subsec:Finite-rotation}

Having studied the effects of mean shear and buoyancy on the flow dynamics, we now consider the effects of rotation. We use the same $Ra=1.25\times10^5$ as in \S \ref{subsec:Zero-rotation} and choose $Pr=5$ to give $Ro = 0.6325$, {which is fixed for all the cases discussed in this section. The range of $Ri_b$ explored here is $Ri_b \in \left[0.1, 100\right]$, slightly larger than in \S\ref{subsec:Zero-rotation}.} 

\begin{figure}
\begin{centering}
\includegraphics[width=0.49\columnwidth]{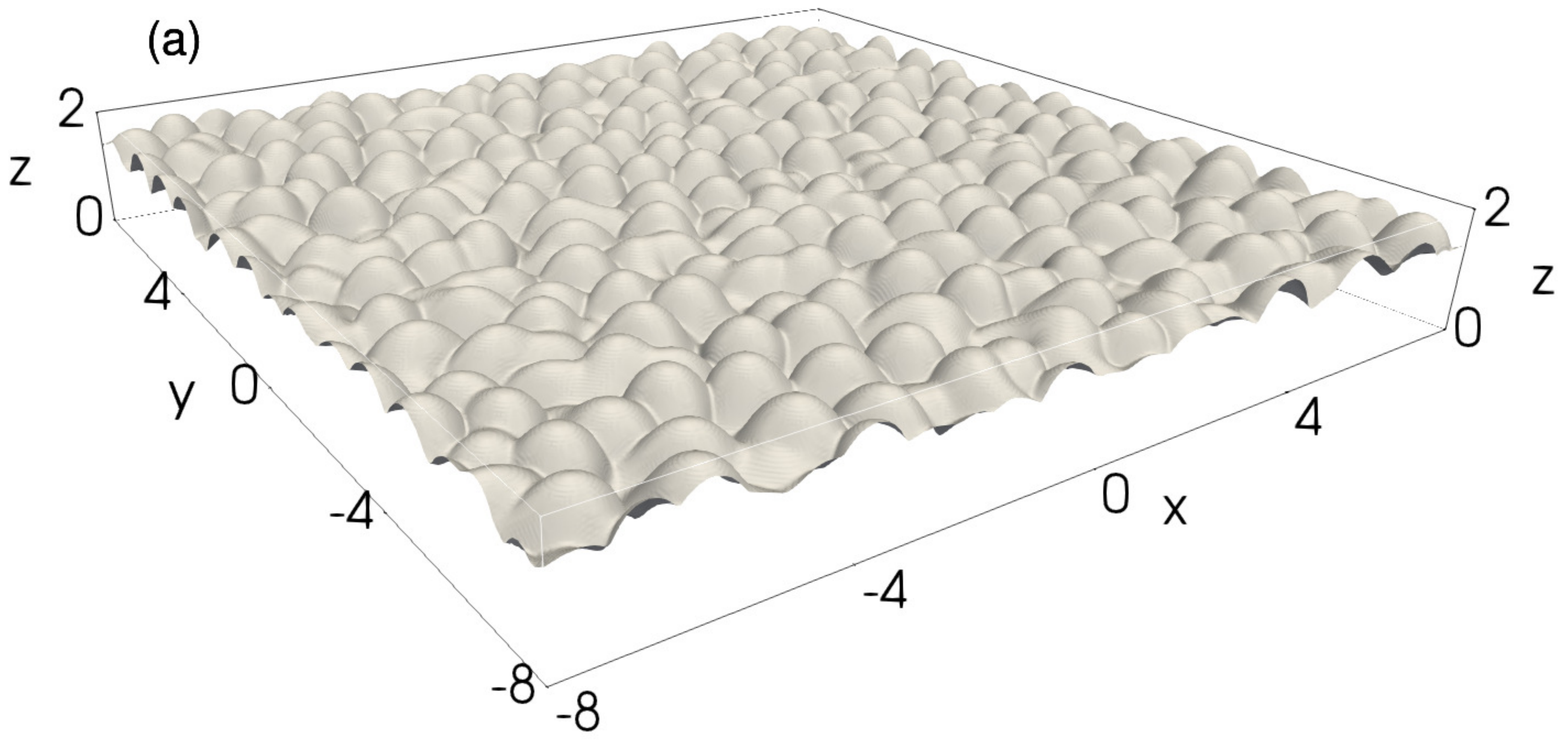}
\includegraphics[width=0.49\columnwidth]{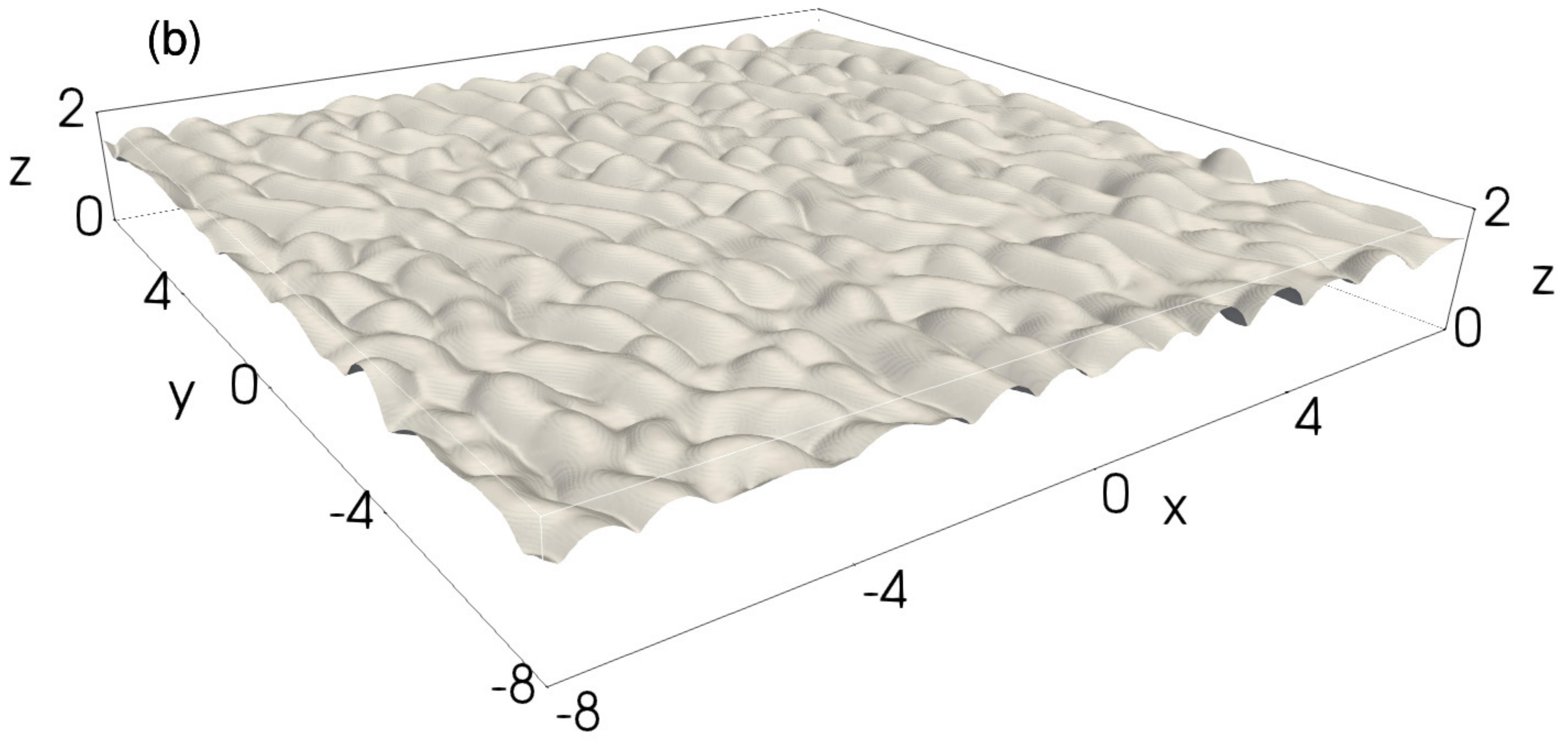}
\par\end{centering}
\begin{centering}
\includegraphics[width=0.49\columnwidth]{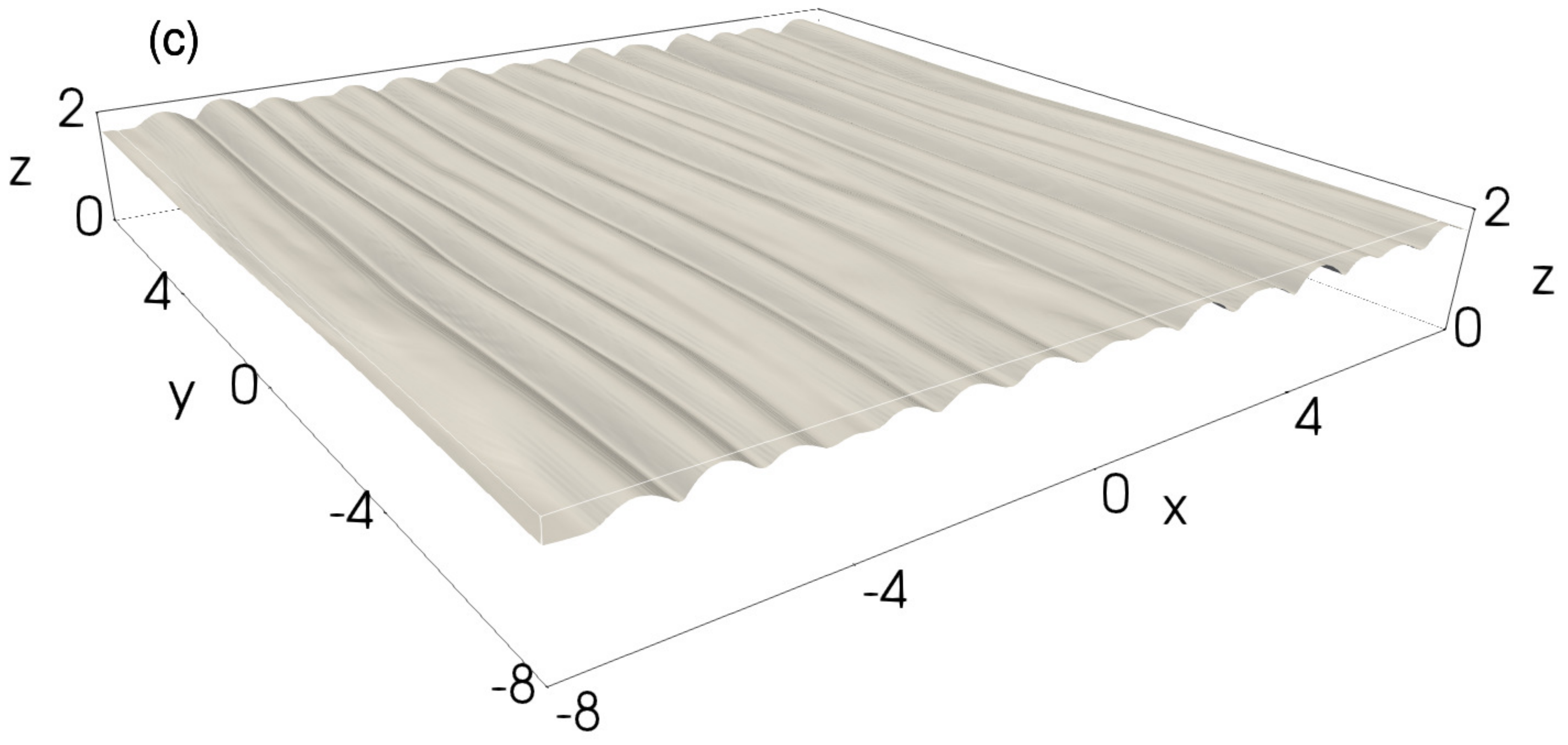}
\includegraphics[width=0.49\columnwidth]{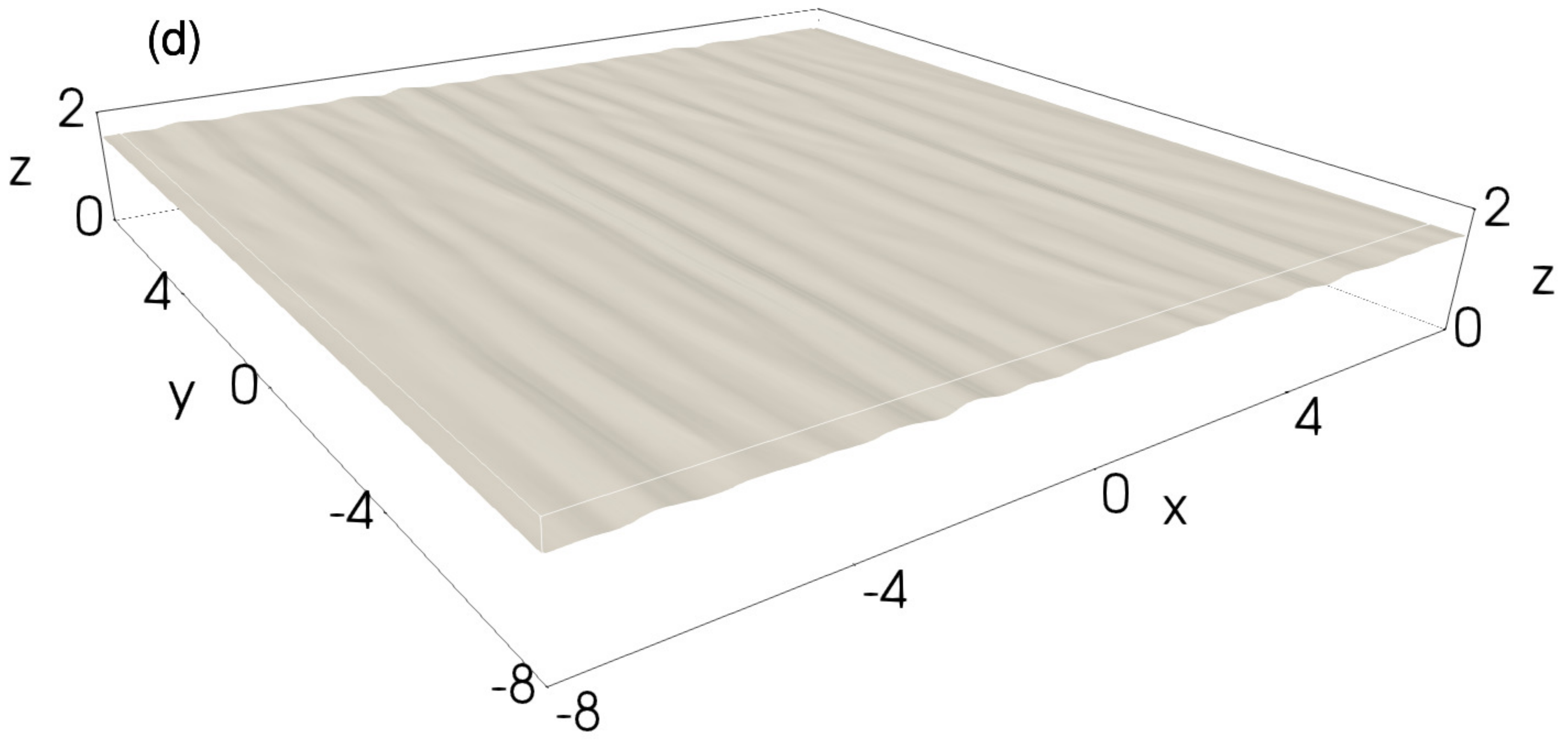}
\par\end{centering}
\caption{\label{fig:solid_liq_interface_E5em4} The solid liquid interface seen from the solid side with flow parameters $Ro = 0.6325$, $Ra=1.25\times10^5$, $Pr=5$, $St=1$, and (a) $Ri_{b}=100$ at $t=155$, (b) $Ri_{b}=10$ at $t=155$, (c) $Ri_{b}=1$ at $t=240$, (d) $Ri_{b}=0.5$ at $t=240$. Characteristic melting by rotating convection \citep[see e.g.][]{ravichandran2021} is seen in (a) for large $Ri_b$. Compare the effect of large $Ri_b$ for $Ro=\infty$ in figure \ref{fig:solid_liq_interface_E0}(b), and note that the figures are plotted at different times owing to the differences in the timescales of melting. (b) For $Ri_{b}=10$, the lateral drift of the vortices is reflected in {the connectivity between neighbouring solid voids} to form longer grooves that do not span the width of the domain. (c) For $Ri_{b}=1$, the grooves become quasi two-dimensional and span the width of the domain in the lateral direction. (d) For $Ri_{b}=0.5$, the grooves are less prominent than in (c). {The times chosen are such that the liquid layer depths are comparable. The values of $\Sigma$ for these cases are (a) 250, (b) 25, (c) 2.5, and (d) 1.25, respectively. Hence, in all these cases, rotation dominates over mean shear.}}
\end{figure}

{Figures \ref{fig:solid_liq_interface_E5em4}(a)-(d) show the phase-boundary geometry for the different $Ri_b$ {when the solid layer has nearly completed melted}. For $Ri_b = 100$, the effects of shear are small and the phase boundary has the regular dome-like features seen in purely rotating case, where columnar vortices span the height of the liquid layer and transport heat between the lower and upper boundaries \citep{rossby1969, ravichandran2021}, as seen in figure \ref{fig:solid_liq_interface_E5em4}(a). 
Upon decreasing $Ri_b$, the homogeneous distribution of domes is replaced by a weak transverse structure, as seen for $Ri_b = 10$ in figure \ref{fig:solid_liq_interface_E5em4}(b). 
With a further decrease in $Ri_b$, the phase boundary develops patterns that are similar to the ones seen in absence of rotation (compare figures \ref{fig:solid_liq_interface_E0}(c), (d) and \ref{fig:solid_liq_interface_E5em4}(c), (d)).  The alignment of the longitudinal corrugations/grooves is differs between rotating and non-rotating cases because in the former case the flow in the bulk is determined by the balance between the Coriolis effect and the applied pressure gradient.  Therefore, the mean shear flow in the bulk is in the negative $y$ direction rather than in the positive $x$ direction.  
Hence, despite the external pressure gradient applied in the positive $x$-direction, the longitudinal grooves in the presence of rotation are parallel to the $y$-axis. }

\begin{figure}
\includegraphics[width=1\columnwidth]{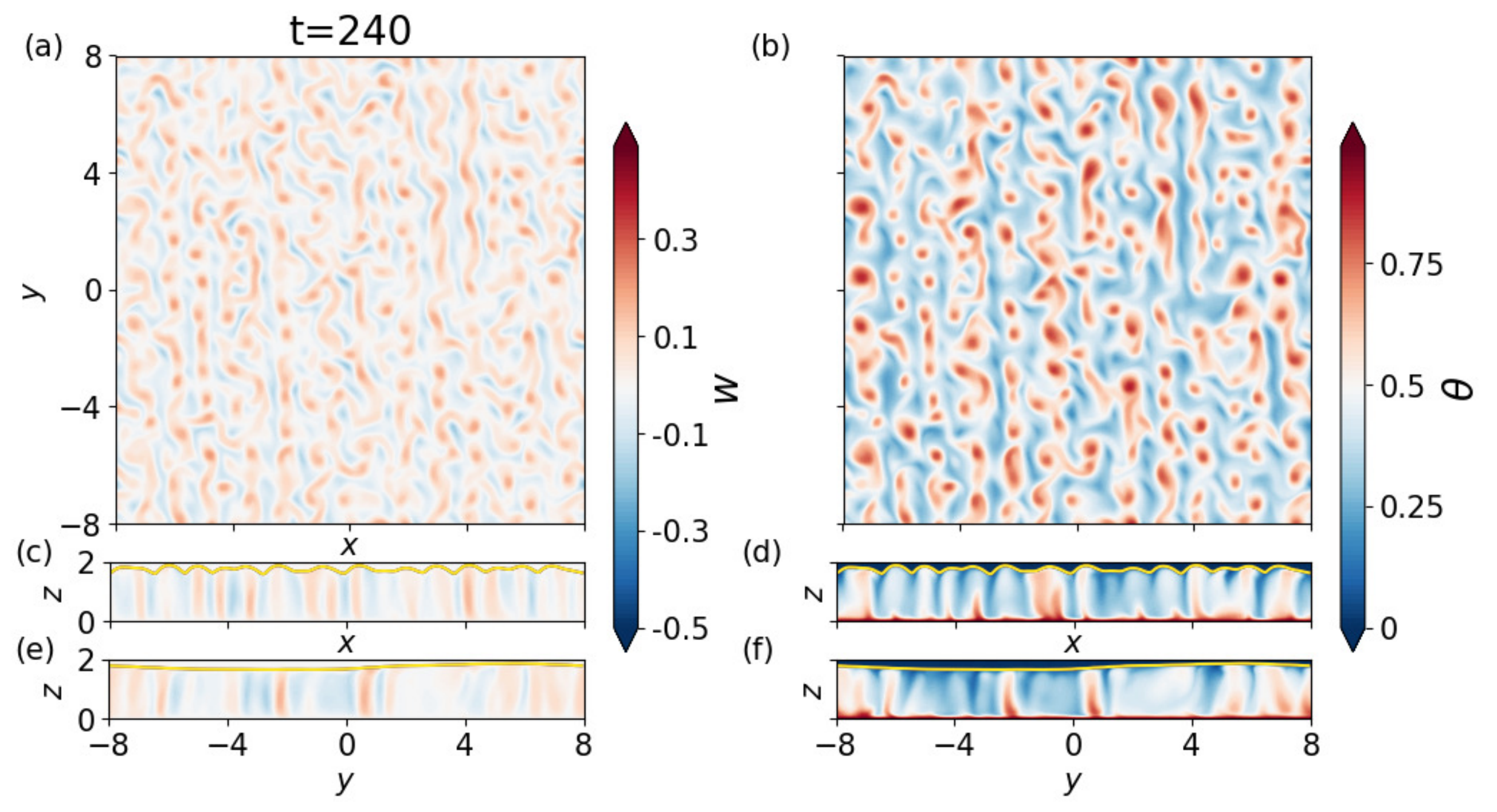}
\caption{\label{fig:Rep1_E5em4} {Contours of vertical velocity $w$ and temperature $\theta$ at the horizontal section $z=H/2$ (a,b) and the vertical sections $y=0$ (c,d) and $x=0$ (e,f). The solid-liquid interface is shown in the vertical sections as a solid yellow line. The flow parameters are $Ri_b=1$, $Ro = 0.6325$, $Ra=1.25\times10^5$, $Pr=5$, and $St=1$. \tb{The value of $\Sigma$ is $0.25$ for this case, so that mean shear dominates rotation.} The drift velocity of the columnar vortices along the negative $y$ direction is sufficiently strong for the melting morphology to become quasi-two dimensional (compare the vertical sections (e) and (f) to the melting morphology shown in figure \ref{fig:solid_liq_interface_E5em4}(c)).}}
\end{figure}

{The merging of the columnar vortices for $Ri_b = 1$ can also be seen in the vertical velocity and temperature fields, as shown in figure \ref{fig:Rep1_E5em4}. In particular, figures \ref{fig:Rep1_E5em4}(a,b) show some of the merger events and the incipient longitudinal streaks that develop as a result of the mean shear flow. These features can also be seen in the vertical cross-sections of the flow (figures \ref{fig:Rep1_E5em4}(c-f)).}

\begin{figure}
\begin{centering}
\includegraphics[width=1\columnwidth]{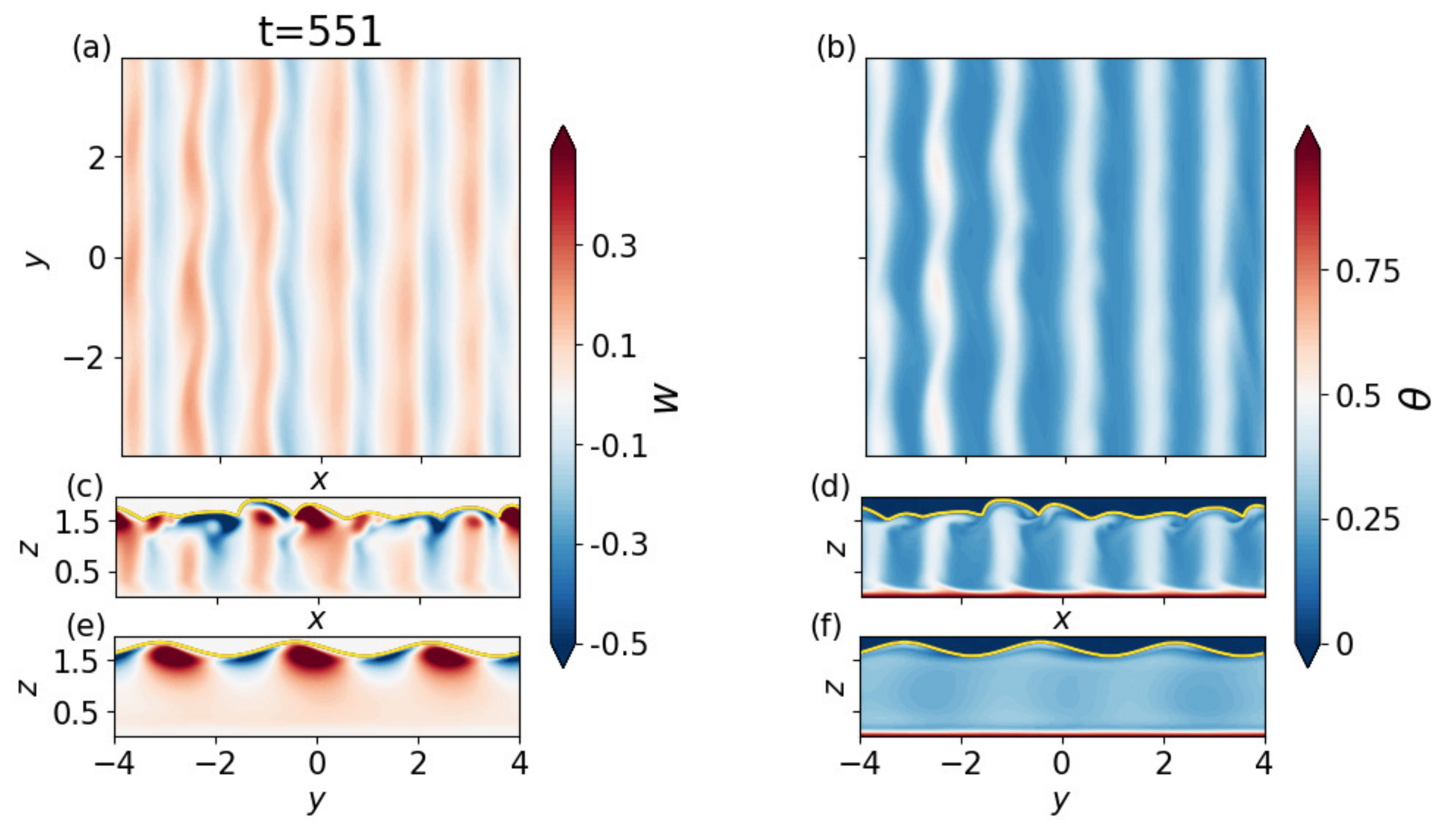}
\par\end{centering}
\caption{\label{fig:wT_Rep10_h0d25} Contours of vertical velocity $w$ and temperature $\theta$ as in figure  \ref{fig:Rep1_E5em4}, and for the same parameters except $Ri_b=0.1$ \tb{($\Sigma=0.25$)} and initial liquid layer height $h_0=0.5$. The columnar vortices are no longer distinguishable and have merged to produce two-dimensional rolls in the bulk (see also figure \ref{fig:uv_mean_Rep10}). The turning of the flow in the Ekman layers creates the grooves aligned at an angle intermediate between the $x$  and $y$ directions.}
\end{figure}

\begin{figure}
\begin{centering}
\includegraphics[width=0.5\columnwidth]{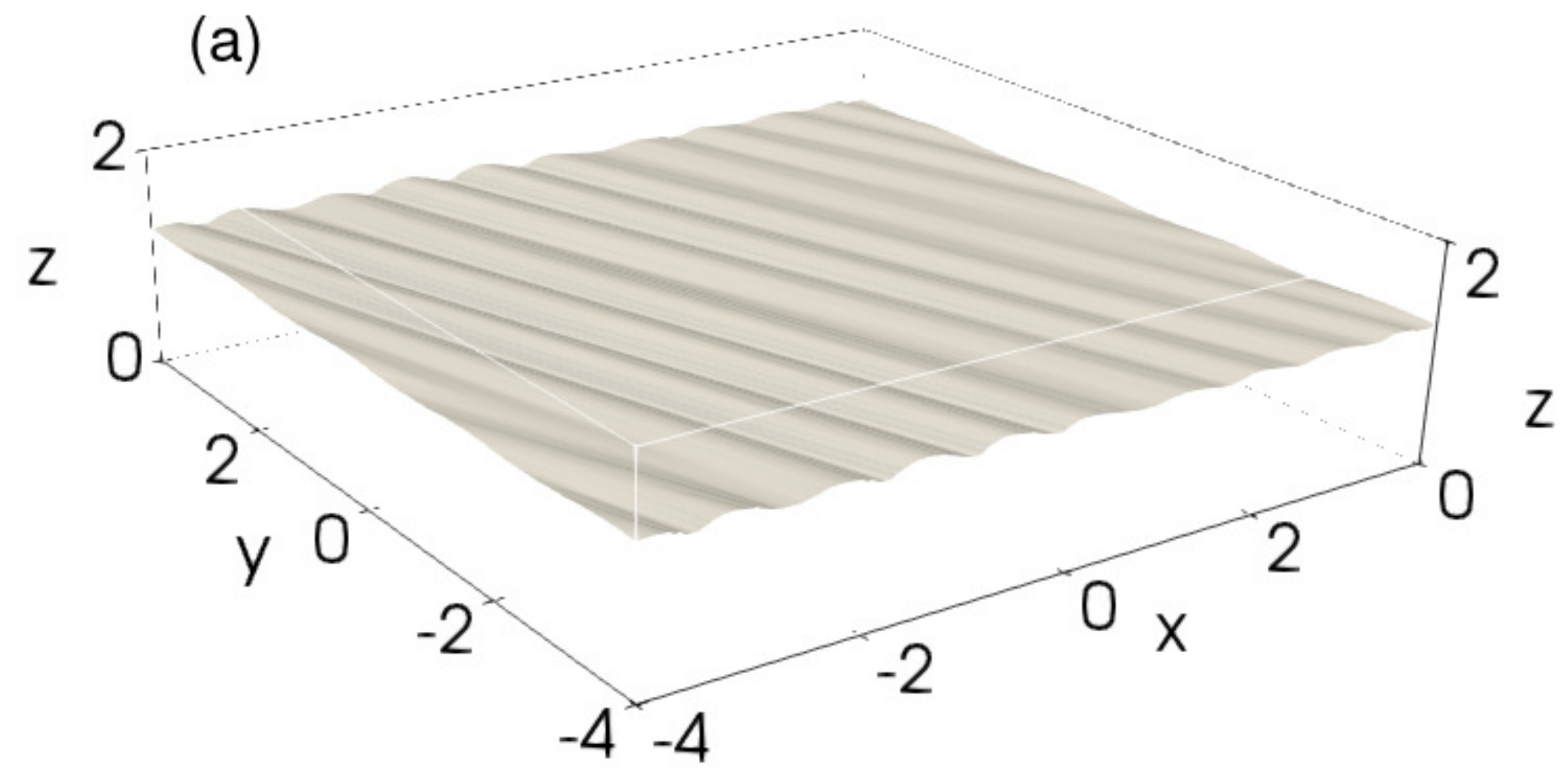}\includegraphics[width=0.5\columnwidth]{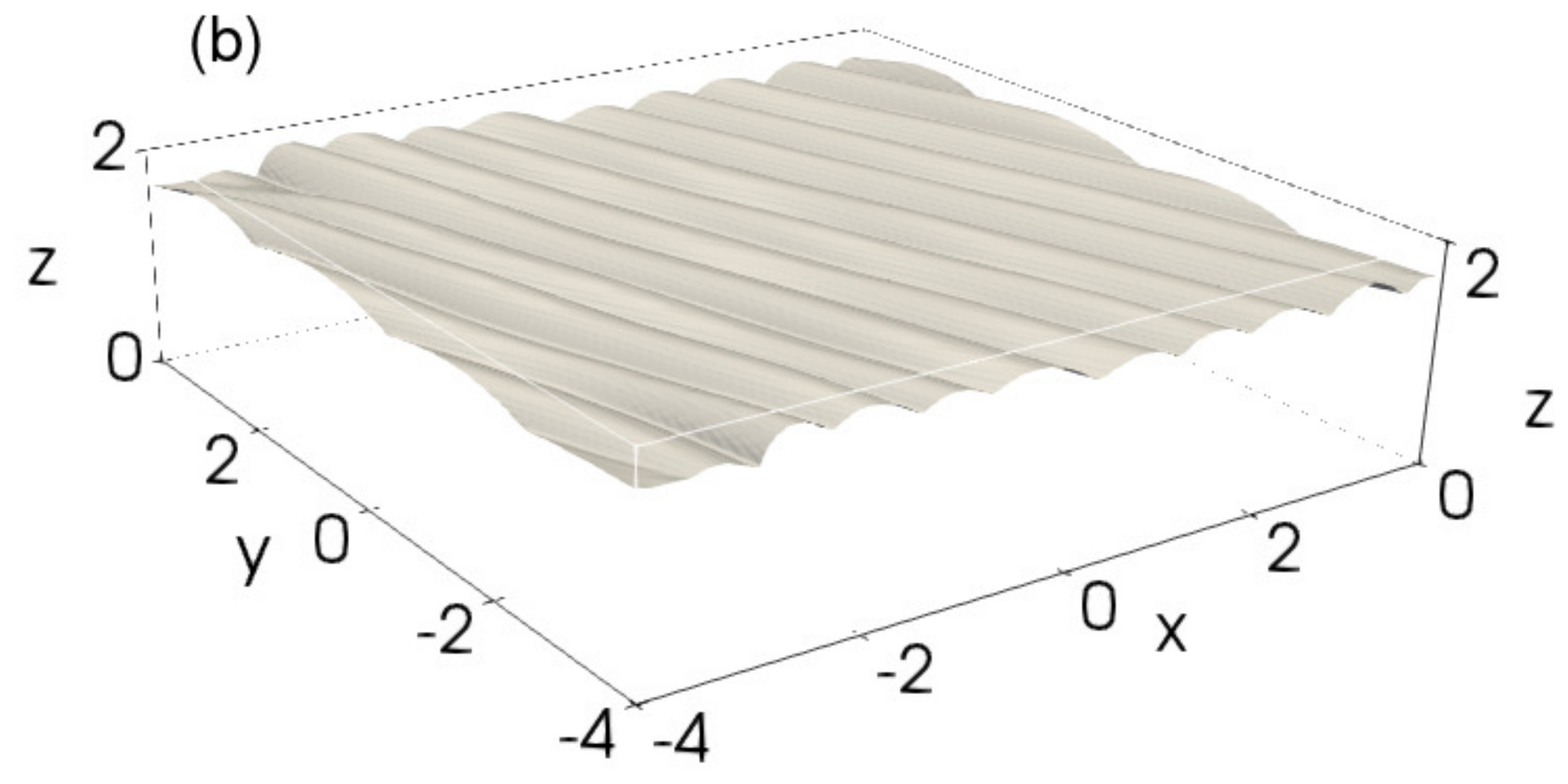}
\par\end{centering}
\begin{centering}
\includegraphics[width=0.5\columnwidth]{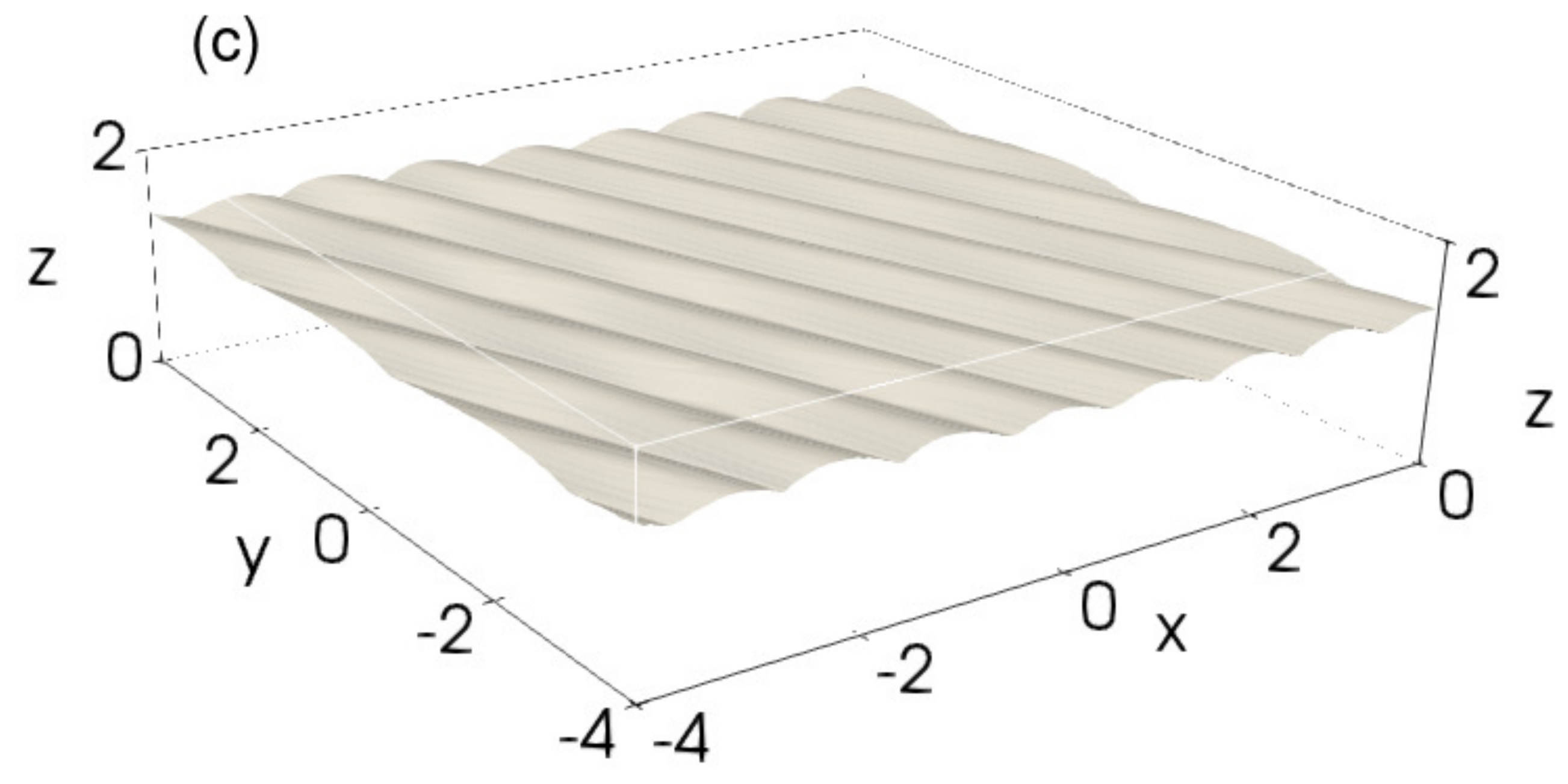}\includegraphics[width=0.5\columnwidth]{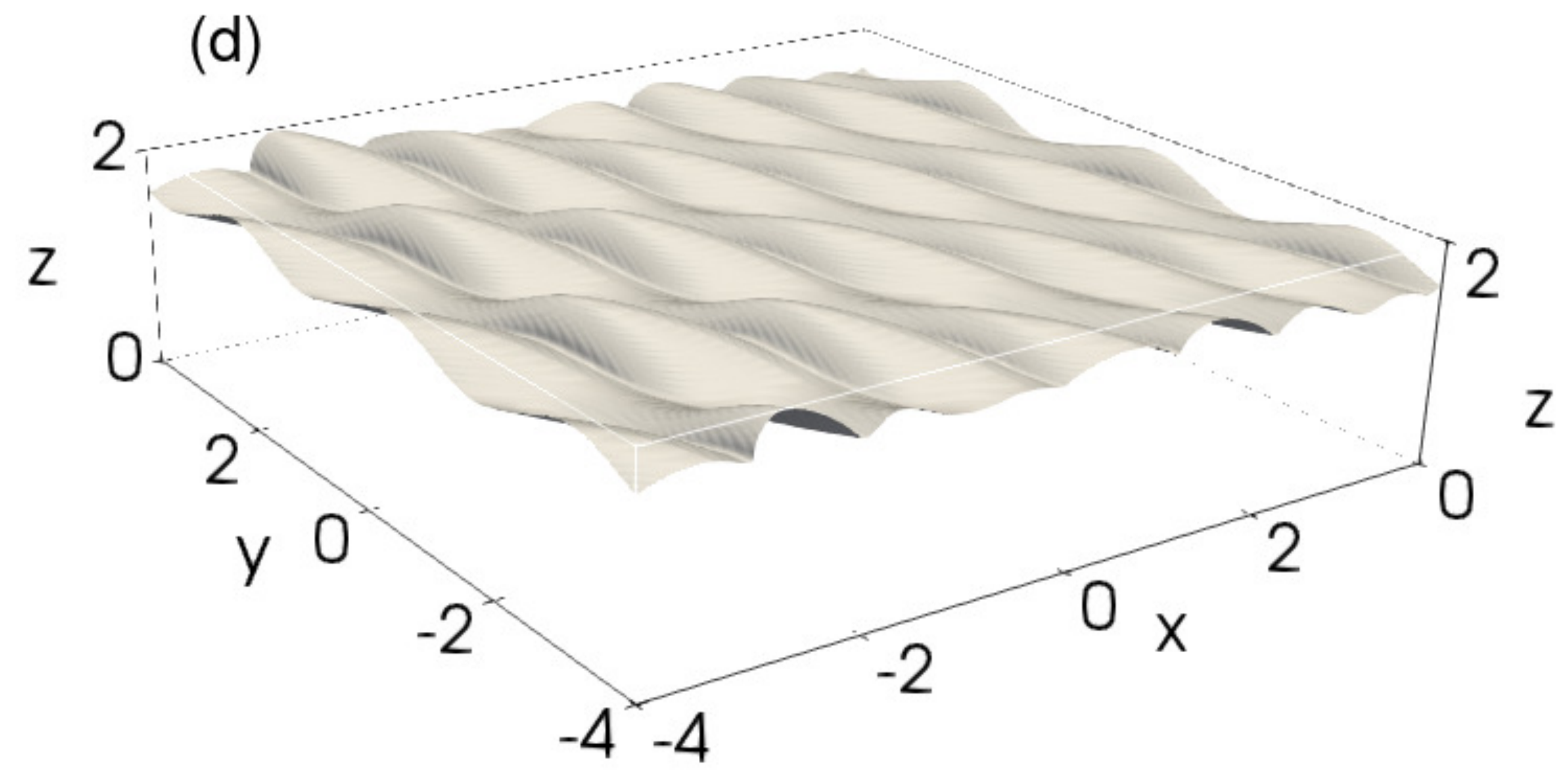}
\par\end{centering}
\caption{\label{fig:solid_liq_interface_Rep5_10_h0d25} The morphology of the solid-liquid interface (seen from the solid side) at (a) $t=354$ and (b) $t=495$ for $Ri_{b}=0.2$, and (c) $t=495$ and (d) $t=552$ for $Ri_{b}=0.1$. {Here, $\Sigma < 1$ for both cases and mean shear also dominates rotation}. As melting proceeds, the grooves are aligned in a direction intermediate between the $x$  and $y$ directions, with a smaller angle between the grooves and the $x$ axis for smaller $Ri_{b}$. At late times, the interface develops a sinusoidal mode {parallel to} the grooves. The initial liquid height $h_{0}=0.5$ in these simulations. The angle between the grooves and the $x$-axis is smaller for smaller $Ri_b$.}
\end{figure}

{The loss of coherence of the columnar vortices is complete for $Ri_b \le 0.2$, and the longitudinal rolls now clearly appear in the flow field. This is seen in figure \ref{fig:wT_Rep10_h0d25}, which shows the contours of vertical velocity and temperature for $Ri_b = 0.1$.} 
{In addition to the complete merger of the vortices, one can also see that the longitudinal rolls meander about their mean location. This indicates the onset of a wave-like instability of the rolls that is qualitatively similar to the one seen in thermal convection with mean shear flows, but without rotation \citep{clever1991, clever1992, Pabiou2005}. These interfering waves are clearly reflected in the patterns at the phase boundary, as seen in figure \ref{fig:solid_liq_interface_Rep5_10_h0d25}(d).}

\begin{figure}
\includegraphics[width=1\columnwidth]{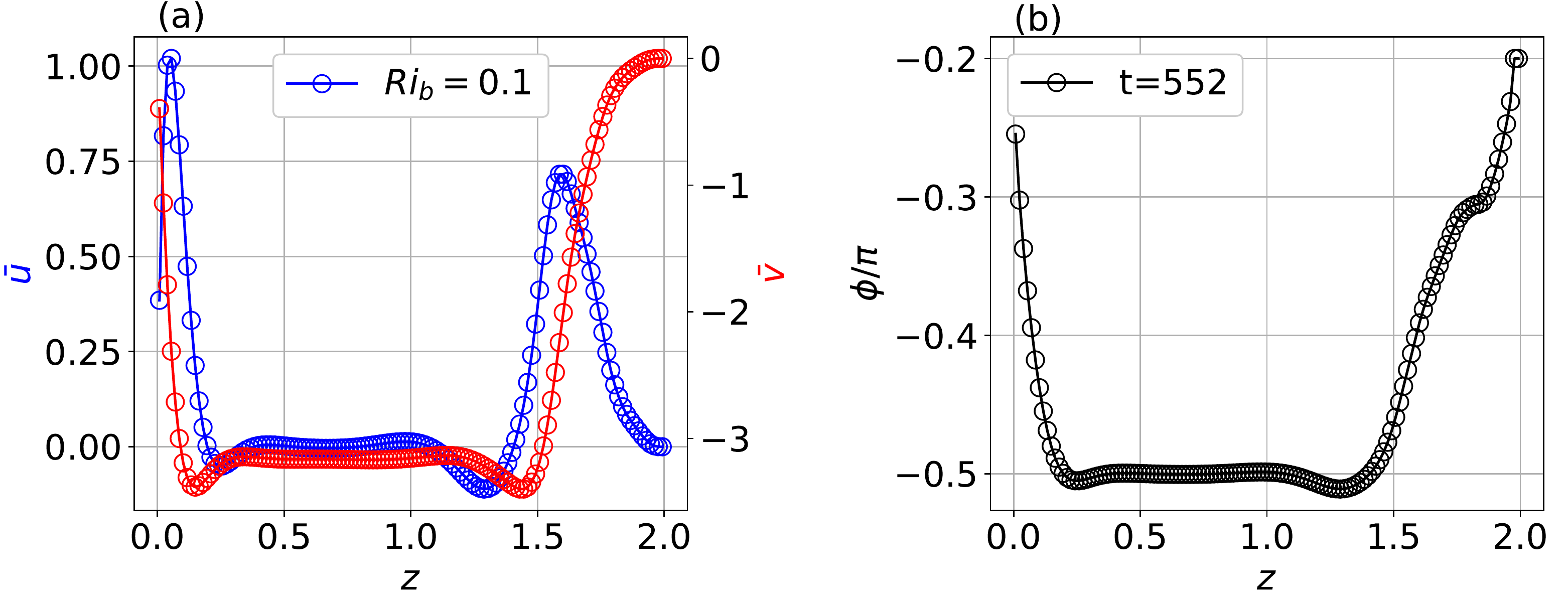}
\caption{\label{fig:uv_mean_Rep10} {(a) The horizontally averaged mean values of the velocities $u$ and $v$ along the $x$  and $y$ directions respectively, and (b) the angle \tb{$\phi=\tan^{-1}(\bar{v}/\bar{u})$} relative to the applied pressure gradient } for $Ri_b=0.1$ at $t=552$  corresponding to the snapshot in figure \ref{fig:solid_liq_interface_Rep5_10_h0d25} (d), as shown by the labels for $Ri_b$ in the inset. The fluid velocity is nearly perpendicular to the applied pressure gradient {in the bulk, and turns counter-clockwise in the Ekman layer at the solid-liquid interface. For $Ri_b/Ro^2<1$, this counter-clockwise turning of the flow in the upper Ekman layer leads to the observed formation of grooves at an oblique angle. Note that the grid point at $z=0$ is not shown.} }
\end{figure}

{To understand the alignment of the longitudinal corrugations with respect to the \tb{applied pressure gradient}, we study the mean flow in the bulk. In figure \ref{fig:uv_mean_Rep10}, we plot the horizontal components of the mean velocity averaged over the horizontal planes and the angle the mean horizontal velocity makes with the applied pressure gradient for $Ri_b=0.1$.} 
{These plots show that despite the strength of the \tb{applied pressure gradient}, the flow in the bulk remains in geostrophic balance (and thus the angle made by the flow relative to the $x$ direction is $\phi=-\pi/2$). However, the no-slip upper and lower boundaries force the mean flow direction to rotate and form Ekman spirals. 
Thus, this counter-clockwise rotation of the mean flow direction ($\phi$ increases from $-\pi/2$ to $0$) in the Ekman layer adjacent to the upper boundary has profound effects on the melting morphology.  This is seen in figure \ref{fig:solid_liq_interface_Rep5_10_h0d25}, which shows the formation of grooves at the phase-boundary for $Ri_b = 0.2$ (figures \ref{fig:solid_liq_interface_Rep5_10_h0d25}(a) and (b)) and $Ri_b = 0.1$ (figures \ref{fig:solid_liq_interface_Rep5_10_h0d25}(c) and (d)). In contrast to the large $Ri_b$ cases, these grooves are neither aligned parallel nor perpendicular to the direction of the applied pressure gradient but at an intermediate angle. 

\tb{The Ekman layer at the upper boundary may be defined as the region where the angle $\phi \neq -\pi/2$ (see figure \ref{fig:uv_mean_Rep10}(b)). For smaller $Ri_b$, the average horizontal velocity in this region is larger and makes a slightly smaller angle with the $x-$axis, decreasing from $85^\circ$ for $Ri_b=0.2$, to $80^\circ$ for $Ri_b=0.1$ (at the times shown in figure \ref{fig:solid_liq_interface_Rep5_10_h0d25}). Similarly, we find that the angle between the grooves and the $x$-axis decreases slightly, from $75^\circ$ to $70^\circ$, as $Ri_b$ decreases. }

{In the absence of rotation (figure \ref{fig:solid_liq_interface_E0}), and with finite background rotation for $Ri_b \geq {O}(1)$ (figure \ref{fig:solid_liq_interface_E5em4}), the grooves formed are stationary in that once formed they do not move, but become more prominent with time (see figure \ref{fig:hovmoller_E0}). In contrast, the obliquely aligned grooves for small $Ri_b$ are advected perpendicular to {their lengths},
as seen from the \Hovmoller plots in figure \ref{fig:hovmoller_E5em4_h0d25} for $Ri_b=0.1$. Since the propagation speeds are proportional to the wavelengths in the $x$ and $y$ directions, figures \ref{fig:hovmoller_E5em4_h0d25}(a,b) also show that the grooves translate horizontally perpendicular to their lengths.} 
\begin{figure}
\includegraphics[width=1\columnwidth]{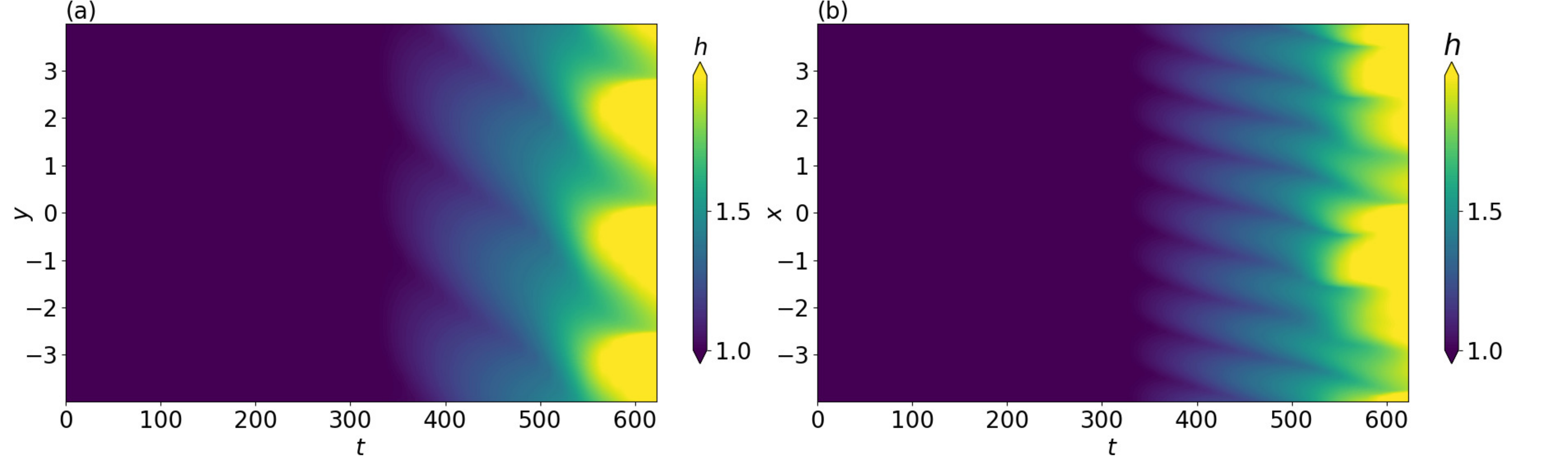}
\caption{\label{fig:hovmoller_E5em4_h0d25} {\Hovmoller plots of the fluid height at (a) $x=0$ and (b) $y=0$, showing the drift of the grooves for the same parameters as in figure \ref{fig:wT_Rep10_h0d25}. The grooves move perpendicular to their lengths 
and thus maintain the same angle to the x-axis for the duration of their existence.}}
\end{figure}

{The impact of the flow and phase-boundary dynamics on the time evolution of $h_s$, $Nu$, and $\sigma_h$ for the cases corresponding to figure \ref{fig:solid_liq_interface_E5em4} are shown in figure \ref{fig:sum_chi_vs_t_E5em4}.}
\begin{figure}
\begin{centering}
\includegraphics[width=1\columnwidth]{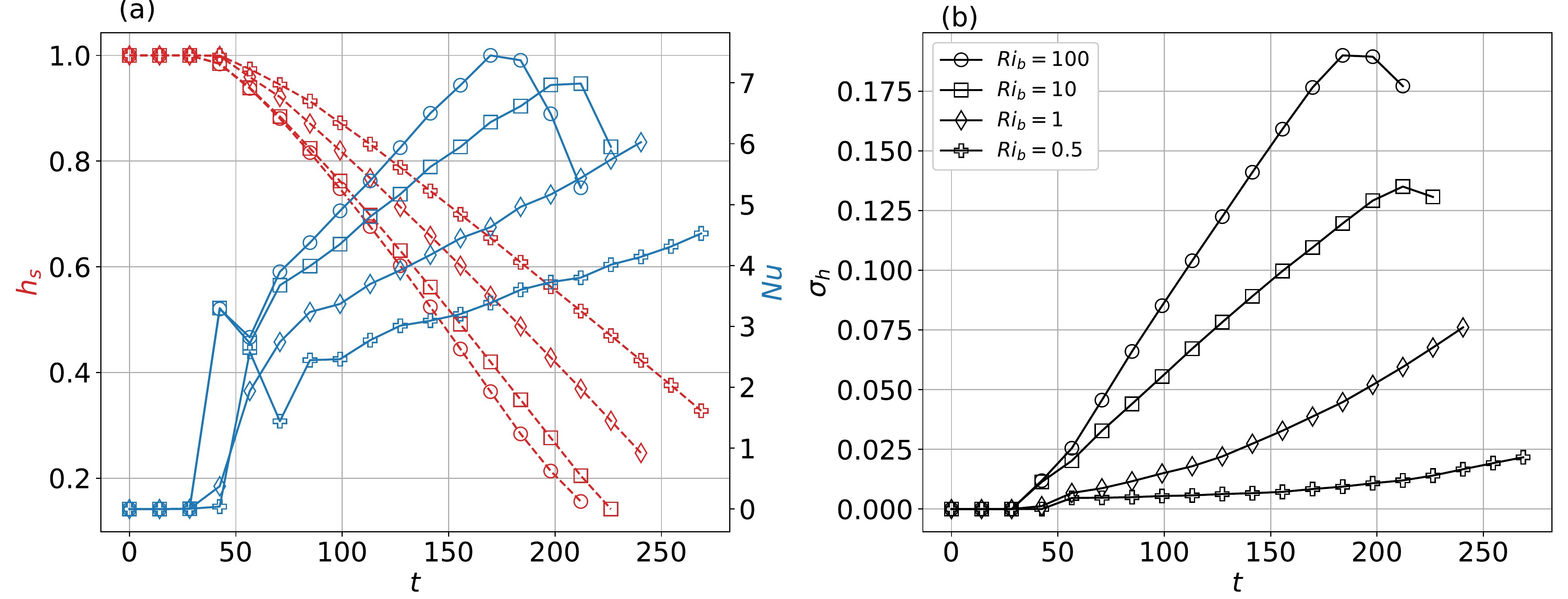}
\par\end{centering}
\caption{\label{fig:sum_chi_vs_t_E5em4} (a) The thickness of the solid layer $h_s$ (red curves, left axis) and the melting Nusselt number (from eq. \ref{eqn:Nusselt}; blue curves, right axis); and (b) the roughness of the solid-liquid interface $\sigma_h$ for $Ri_b=100,10,1,0.5$ \tb{(with $\Sigma= 250, 25, 2.5, 1.25$ respectively)} and other parameters as in figure \ref{fig:Rep1_E5em4}. The rate of melting decreases with decreasing $Ri_{b}$, unlike in figure \ref{fig:sum_chi_vs_t_E0}. We also note that the maxima of $Nu$ and $\sigma_h$ are reached simultaneously for $Ri_b=10$.}
\end{figure}
{The melt rate and the Nusselt number decrease as $Ri_b$ decreases, because as the strength of the shear flow increases vertical convective motions are suppressed, as shown in figure \ref{fig:sum_chi_vs_t_E5em4}(a).  For large $Ri_b$, columnar convection dominates the flow leading to a more ramified dome structure of the phase boundary, which gives way to an aligned periodic phase boundary as $Ri_b$ decreases and shear dominates.  Thus, $\sigma_h$ increases with $Ri_b$, as shown in figure \ref{fig:sum_chi_vs_t_E5em4}(b). 
Note that for $Ri_b = 10$ and $100$, $\sigma_h$ reaches a maximum at the same time as the corresponding $Nu$, as observed in the case of $Ri_b = \infty$ \citep{ravichandran2021}.}

{Similar behaviour in the evolution of $h_s$, $Nu$, and $\sigma_h$ is observed for the smallest three $Ri_b$ cases studied, as shown in figure \ref{fig:sum_chi_vs_t_E5em4_h0d25}.}
\begin{figure}
\begin{centering}
\includegraphics[width=1\columnwidth]{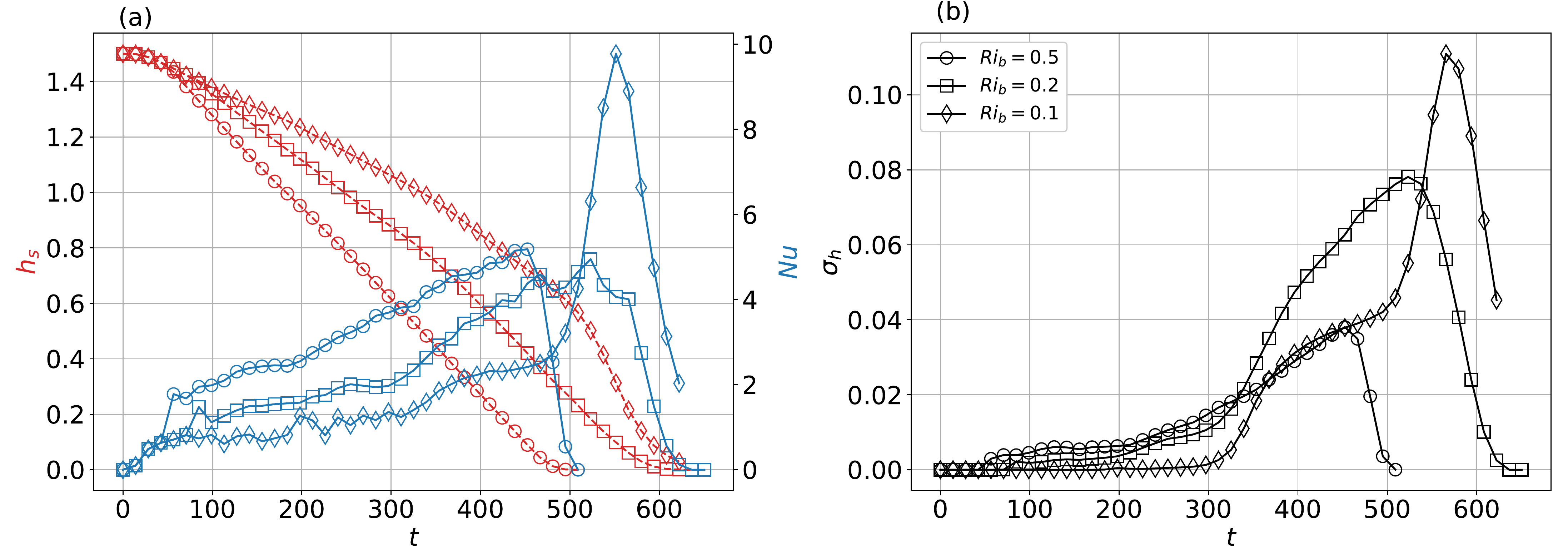}
\par\end{centering}
\caption{\label{fig:sum_chi_vs_t_E5em4_h0d25}  {(a) The thickness of the solid layer and (b) the Nusselt number as a function of time} for $Ro = 0.6325$, $Ra=1.25\times10^5$, $Pr=5$, $St=1$, \tb{with $\Sigma=1.25, 0.5, 0.25$ for $Ri_b=0.5,0.2,0.1$ respectively}. The overall rate of melting is initially smaller for smaller $Ri_{b}$, but as the flow develops, the Nusselt number increases sharply with the development of the along-groove sinusoidal mode. Also note the abrupt change in the slope of the thickness curve for $Ri_b=0.1$ at $t \approx 552$, which coincides with the increase in $Nu$.  }
\end{figure}
{However, for large times, there are substantial jumps in $Nu$ and $\sigma_h$ for $Ri_b = 0.1$ and $0.2$. This is due to the onset of the wave-like interfacial instability of the longitudinal rolls shown in figures  \ref{fig:wT_Rep10_h0d25} and \ref{fig:solid_liq_interface_Rep5_10_h0d25}. This change in the phase boundary geometry leads to the generation of more intense downwelling plumes that result in enhanced heat transfer \citep{TSW2015_EPL}. Another effect of these cold plumes is that the bulk fluid is at a temperature lower than the mean of the top and bottom surfaces as shown in figure \ref{fig:u_theta_vs_z_rot_h0d25} for $Ri_b = 0.1$.}

{In geostrophic convection, the interaction between the thermal and momentum boundary layers plays an important role \citep{rossby1969,julien2012,King2012}. In figure \ref{fig:u_theta_vs_z_rot_h0d25}, we plot the vertical profiles of the horizontally averaged temperature $\bar{\theta}$ and the horizontally averaged velocity $-\bar{v}$ for $Ri_b=0.1$. At early times, the thermal boundary layer is much thicker than the Ekman layer and the heat transport is controlled by rotation. This also results in a significant bulk temperature gradient. As the depth of the liquid layer increases with time, the effective Rayleigh number increases and the thermal boundary layer becomes thinner, becoming comparable to the Ekman layer thickness at $t\approx500$. The crossing of the thermal and velocity boundary layers leads to convection-dominated dynamics, which is evident from the small temperature gradient in the bulk at $t\approx550$ shown in figure \ref{fig:u_theta_vs_z_rot_h0d25}(b).}
\begin{figure}
\begin{centering}
\includegraphics[width=1\columnwidth]{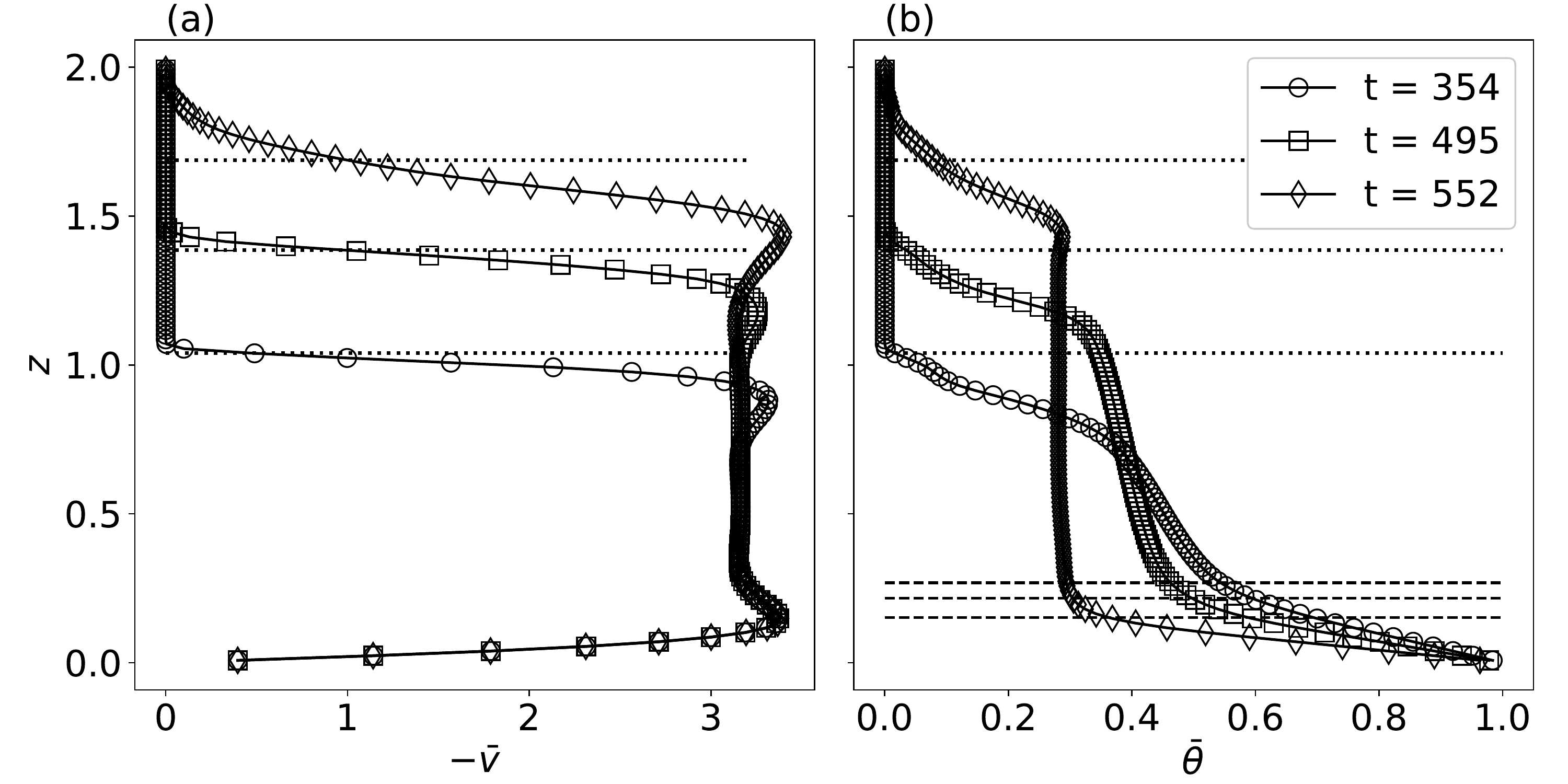}
\par\end{centering}
\caption{{\label{fig:u_theta_vs_z_rot_h0d25} Profiles of the horizontally averaged (a) horizontal velocity $-\bar{v}$ and (b) temperature $\bar{\theta}$ for $Ro=0.6325$, $Ra=1.25\times10^5$, $Pr=5$, $St=1$, $Ri_b=0.1$ at the times shown. The temperature in the fluid bulk has a pronounced temperature gradient due to the combined influence of shear and rotation in the initial stages of evolution; the emergence of the along-groove sinusoidal mode coincides with the disappearance of the bulk temperature gradient. \tb{The dashed lines show the thermal boundary layer at the lower boundary, while the dotted lines show the average height of the fluid layer.}}}
\end{figure}

The different phase-boundary geometries suggest the existence of four distinct regimes{: (1)} $Ri_b \gg 1$, the mean shear flow is weak and rotation and buoyancy dominate, leading to columnar vortices and dome-like features at the phase boundary{; (2)} $Ri_b = O(1)$, the mean shear flow is stronger and the columnar vortices are advected with the flow. This is reflected in the phase boundary geometry which shows incipient grooves perpendicular to the direction of {the applied pressure gradient} (figures \ref{fig:solid_liq_interface_E5em4}(b,c) and \ref{fig:Rep1_E5em4}(a,b)); (3) $Ri_b \lesssim 0.5$, strong shear leads to the loss of coherence of the columnar vortices, resulting in their partial merger, while the interfacial grooves become less prominent. This is seen in figure \ref{fig:solid_liq_interface_E5em4}(d){; (4)} $Ri_b \lesssim 0.2$, the columnar vortices merge completely and convective motions become quasi-two dimensional in the bulk. The flow in these cells is turned counter-clockwise in the Ekman layers, leading to the formation of grooves aligned at an angle to the direction of {the bulk shear flow} and thereby becoming sinusoidal.

\section{Conclusions} \label{sec:Conclusions}

We have systematically explored the effects of buoyancy, rotation, and shear on the evolution of a phase boundary using direct numerical simulations in three dimensions \tb{for} (a) $Ro = \infty, 1$; (b) $Ra=1.25\times10^5$; (c) $St = 1$; (e) $Pr = 1, 5$; and (f) $Ri_b \in \left[0.1, \infty\right)$. The main conclusions from our study are as follows.

(1) In the absence of rotation ($Ro = \infty$), we observe either dome-like features or grooves on the phase boundary depending on whether buoyancy or mean shear dominates {the flow.  In particular, we find the following three features.}
\begin{enumerate}
 \item As the value of $Ri_b$ decreases from $\infty$, the strength of the shear flow increases and both the flow structure and phase boundary geometry transition from {being} three dimensional to quasi-two-dimensional. This is seen in figures \ref{fig:solid_liq_interface_E0}, and \ref{fig:Rep0d1_E0}(a,b).
 
 \item For small $Ri_b$, the grooves formed on the phase boundary are aligned parallel to the direction of the shear flow. {These grooves are due to longitudinal rolls, which are the preferred form of convection in this case \citep{clever1991}.} This is in agreement with the numerical results of \citet{couston2020}, but in contrast to the experimental results of \citet{Gilpin1980}. In our simulations, the bulk $Re = O(10^2-10^3)$, whereas in the experiments of \citet{Gilpin1980} the $Re$ based on the boundary-layer thickness is $O(10^4)$, so that the flow regimes are not directly comparable.  
 
 \item \tb{The dimensionless heat flux, $Nu$,}  is a non-monotonic function of $Ri_b$. For $Ri_b \in [10, 100]$, the mean shear is weaker than buoyancy, but nonetheless inhibits vertical motions, leading to smaller melt rates and vertical heat transport. On further decreasing $Ri_b$ to $1$, the flow becomes three-dimensional and enters the forced turbulent convection regime. This results in larger heat transport and hence an increased melt rate.
\end{enumerate}

(2) When rotation is introduced ($Ro=0.6325$), the mean flow in the interior is in geostrophic balance and thus in a direction perpendicular to the direction of the applied pressure gradient. \tb{The dynamics is governed by the balance between mean shear and rotation, and is represented by the parameter $\Sigma$ (equation \ref{eq:Sigma}), with the imposed shear becoming dominant for $\Sigma<1$.}
\begin{enumerate}
 \item For $Ri_b = 100$ \tb{($\Sigma \gg 1$)}, the shear flow is weak and the phase-boundary consists of dome-like features.
 \item On reducing $Ri_b$ to $10$ \tb{($\Sigma = 25$)}, these dome-like features begin to merge. A further reduction in $Ri_b$ leads to the formation of grooves at the phase boundary.
 \item For $Ri_b \in \left[0.2, 0.5\right]$  \tb{($\Sigma \in \left[ 0.5, 1.25 \right]$)}, the merger is of the dome-like features is complete. The rolls develop a wave-like instability for $Ri_b = 0.2$ and $0.1$  \tb{($\Sigma = 0.5$ and $0.25$, respectively)}, which leads to sinusoidal evolution of the phase boundary. This instability is qualitatively similar to that observed in thermal convection in the presence of a mean shear flow \citep{clever1991, clever1992, Pabiou2005}.
\end{enumerate}

\tb{Our study provides foundational understanding of 
the effects of buoyancy, mean shear and rotation on phase-changing boundaries.  Although we have treated values of $Ra$ and $Re$ that are much larger than previously examined, the parameter range explored here is representative of many but certainly not all geophysical settings \cite[See e.g.,][and refs. therein]{meakin2009, Neufeld:2010, bushuk2019, Weady:2022}.}  
Our results suggest several avenues for further research. With or without rotation, our findings suggest that a secondary instability disrupts the basic flow state in which the applied pressure gradient is balanced by other forces in the flow. The mechanism of the instability, and the boundary between the quasi-two dimensional state and the three dimensional state are {compelling foci for further study}. In particular, in rotating convection, the competition between the geostrophic bulk and the viscous Ekman layers is crucial in determining the state of the flow and the concomitant heat transport, which is a question of significant interest in the literature \citep{King2012, julien2012}. Here, however, this competition is reflected in the morphology of the solid-liquid interface, with either the bulk flow or the Ekman boundary layers controlling the alignment of the observed grooves. Therefore, the manner in which the free boundary reflects the state of the flow and the associated flow-geometry feedback, provide an ideal testing ground for general questions in rotating convection and a framework for the wide range of settings, from geophysical to industrial, in which such flows accompany phase changes.


\section*{Acknowledgements}
Computational resources from the Swedish National Infrastructure for Computing (SNIC) under grants SNIC/2019-3-386 and SNIC/2020-5-471 are gratefully acknowledged. Computations were performed on Tetralith. S.R. and J.S.W. acknowledge support from Swedish Research Council under grant no. 638-2013-9243, and S.T. acknowledges a Research Fellowship from All Souls College, Oxford. Nordita is partially supported by Nordforsk.

\section*{Declaration of interests}
The authors report no conflict of interest.


%

\end{document}